\documentclass[review]{elsarticle}

\usepackage{amsmath}
\usepackage{amsfonts}
\usepackage{mathrsfs}
\usepackage{dsfont}
\usepackage{bm}
\usepackage{cases}
\usepackage{algorithm}
\usepackage{algorithmic}
\usepackage{array}
\usepackage{graphicx}
\usepackage{subfigure}
\usepackage{geometry}
\usepackage{float}
\usepackage{epstopdf}
\usepackage{subeqnarray}
\usepackage{amssymb}
\usepackage{bbm}
\usepackage{booktabs}
\biboptions{numbers,sort&compress}

\usepackage[figuresright]{rotating}
\usepackage{caption}

\usepackage{color}  
\definecolor{agreen}{rgb}{0,0.5,0}

\journal{Simulation Modelling Practice and Theory}

\begin{document}
\captionsetup[figure]{labelfont={bf},labelformat={default},labelsep=period,name={Fig.}}
\begin{frontmatter}

\title{A Simulation Method for MMW Radar Sensing in Traffic Intersection Based on BART Algorithm}

\author{Ming~Zong~and~Zhanyu~Zhu\corref{mycorrespondingauthor}}
\cortext[mycorrespondingauthor]{Corresponding author. E-mail address: zyzhu@suda.edu.cn}
\address{School of Electronic and Information Engineering, Soochow University, Suzhou 215006, China}

\begin{abstract}
Millimeter-wave (mmw) radar is indispensable for Intelligent Transportation Systems (ITS), which can monitor traffic conditions in all weathers. An end-to-end simulation method for mmw radar monitoring and identification at traffic intersections is proposed in this paper. In this method, a virtual intersection scenario model is constructed, and the scattering coefficient of the target is calculated using the Bidirectional Analytical Ray Tracing (BART) algorithm. Combined with the generation of time-domain waveforms, the operation of frequency-domain convolution is simplified by inverse Fourier transform, and the echo signals received by the sparse array are simulated. After raw signal processing, point cloud images containing target position information and Range-Doppler Map (RDM) containing target state feature are obtained. The performance of mmw radar in detecting the specific location information of the target is evaluated by analyzing point cloud images. In addition, a self-defined convolutional neural network is introduced in this paper to evaluate the object recognition performance of the RDM. After the training of the neural network, the classification accuracy of this method for four types of vehicle targets can reach 92\%.
\end{abstract}


\end{frontmatter}


\section{Introduction}
In the field of intelligent transportation, mmw sensing has distinct advantages. In contrast to infrared, lidar, and cameras, mmw radar does not rely on weather or light conditions for continuous detection throughout the day and night \cite{1}\cite{7870764}. The information collected by mmw radar installed at traffic intersections is primarily about distance, speed, angle, trajectory, traffic flow, queue length, etc. of vehicles, whose processing is complicated and time-consuming. In the process of system design, simulation techniques can help to shorten the design cycle and save the development costs of the mmw radar system.

Several radar simulators have been used in the verification and testing of vehicle-mounted radars \cite{2}\cite{3}, but few studies have focused on their application to the experimental debugging and performance evaluation of roadside radars. Presently, the division of the radar frequency bands for traffic intersection applications is not clear, which limits deployments of related systems and scenarios to early internal tests and demonstrations. Using the electromagnetic computing theory to perform mmw radar simulation, it is possible to quickly obtain the detection parameters and identify the system performance, which shows great significance for system evaluation, verification, and feasibility analysis in the early stages of the system design. In this paper, the electromagnetic calculation theory is used to simulate the millimeter wave radar. Combined with the whole process of physical detection, such as echo generation, signal processing, etc., the detection parameters can be obtained and the system performance can be identified. It is useful for early system evaluation, verification and feasibility analysis of system design important meaning. In practical applications, the usage of the radar simulator can reduce the period of actual experimental measurement and ensure the experiment safety \cite{4}\cite{5}\cite{6}. The parameters of the radar sensor and the type of traffic vehicles can be easily changed in the simulated traffic scenario at any time, thus providing abundant experimental basic data for subsequent experimental research, such as vehicle recognition and classification \cite{7}\cite{8}, vehicle trajectory tracking \cite{9}\cite{10}\cite{11} and so on.

Different from the traditional 1D range image or 2D image detection methods, the current application of mmw radar in the field of transportation expects to obtain more dimensional target information. For example, the 4D mmw radar applied in the ADAS (Advanced Driving Assistance System) \cite{12} can obtain 3D position and 1D velocity information of the target and then form visual point cloud data. Traditional radar simulators usually can only simulate 1D envelopes of moving targets or 2D images of stationary targets \cite{13}, but lack of the ability to simulate high-dimensional target information. Radar target simulation of high-frequency band is often accelerated by Ray Tracing (RT) technology \cite{14}. However, the discrete sampling of RT technology leads to incoherent scattering computing when the incident angle changes and the generated data is difficult to analyze further using coherent processing.

This paper proposes an end-to-end mmw radar detection and recognition simulation method. The Bidirectional Analytical Ray Tracing (BART) algorithm is used to calculate the Radar Cross-Section (RCS) of the target, and the speed of target simulation calculation is accelerated. Combined with the transmitted signal generated in the time domain, the echo signal can be calculated, and the signal processing process can be realized digitally. All intermediate processes and results can be controlled, which is convenient for analyzing the whole process. This method can output the point cloud and RDM of the simulated target, including the position information and state information of the target respectively. Based on this simulation method, this paper also uses the deep learning method to conduct target classification experiments on the simulated data \cite{9568916}\cite{21061951}, and the performance and feasibility of the application of millimeter-wave radar at traffic intersections are analyzed. The rest of this paper is organized as follows: Section 2 briefly introduces the basic principles of mmw radar detection and the scattering characteristics of the target. Section 3 introduces the construction of the intersection simulation scenario, the simulation processing of the mmw radar signal, the data set of the vehicle classification network and the convolutional neural network used. Section 4 introduces the specific operations of the simulation experiments, including the detection and classification of vehicle objects, and analyzes the experimental results. Section 5 summarizes the work of the full paper.

\section{Related technologies}
This paper mainly involves the principle of radar point cloud imaging and the principle of RCS computing based on BART, which will be briefly introduced in this section.
\subsection{Target detection principle of mmw radar}
\begin{figure*}[!t]
	\centerline
	{
		\includegraphics[width=12cm]{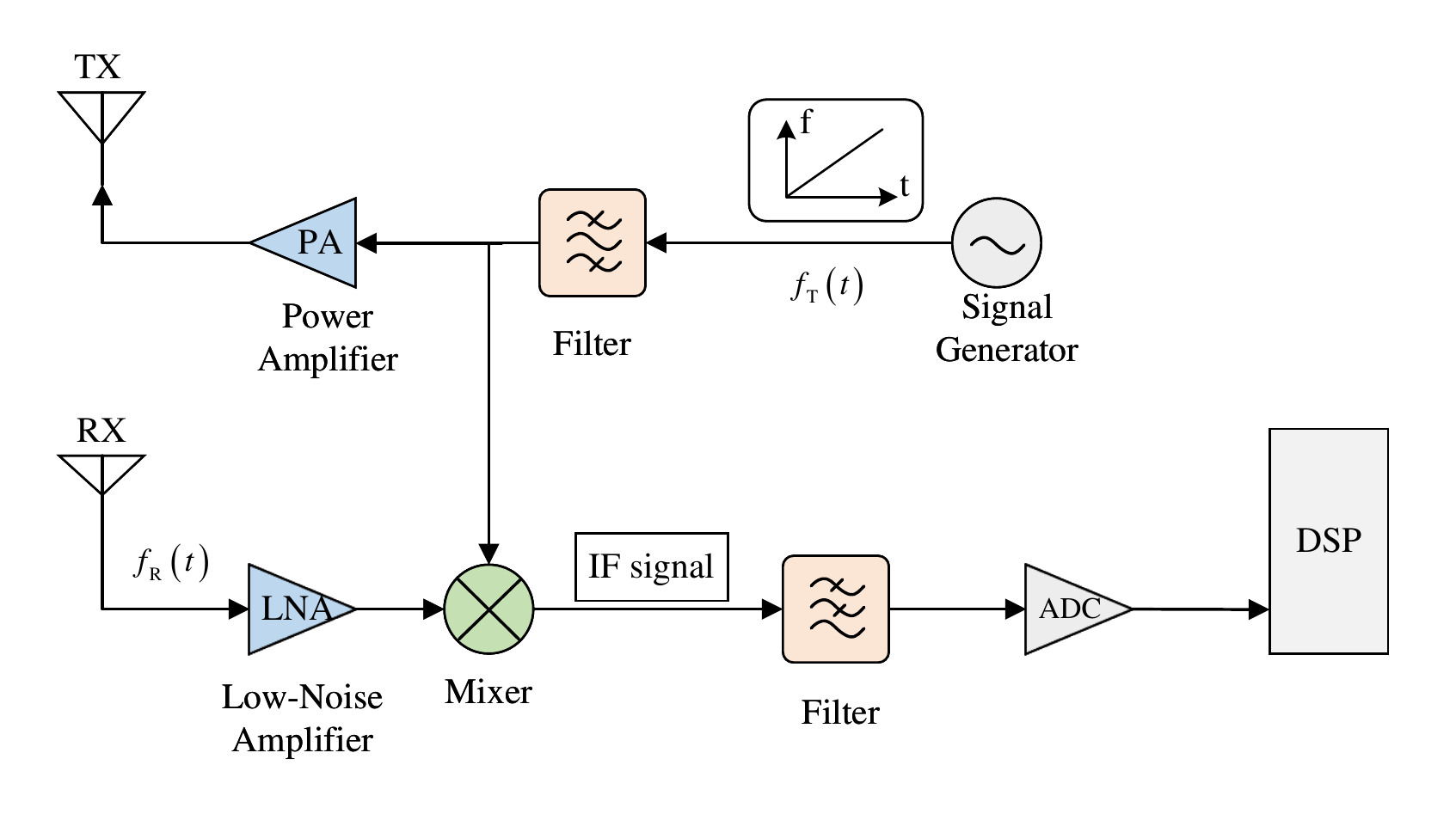}
	}
	\caption{Mmw radar system block diagram.}
	\label{Fig1}
\end{figure*}

Because of the simplicity of the design of linear Frequency-Modulated Continuous-Wave (FMCW) radar systems, it is often chosen as the transmit signal in near-field applications \cite{8828025}. This paper mainly studies the mmw radar system based on FMCW, and its system block diagram is shown in Fig. \ref{Fig1}. The signal generator generates the signal ${{\it{f}}_\text{T}}(t)$ and transmits it through the antenna TX. When the signal touches the target, part of the signal will be captured by the receiving antenna RX after reflection, and received as a radar echo signal ${{\it{f}}_\text{R}}(t)$ \cite{15}. Target features such as distance, speed, and angle can be obtained after a series of data processing on ${{\it{f}}_\text{R}}(t)$.

\begin{figure*}[!t]
	\centerline
	{
		\includegraphics[width=12cm]{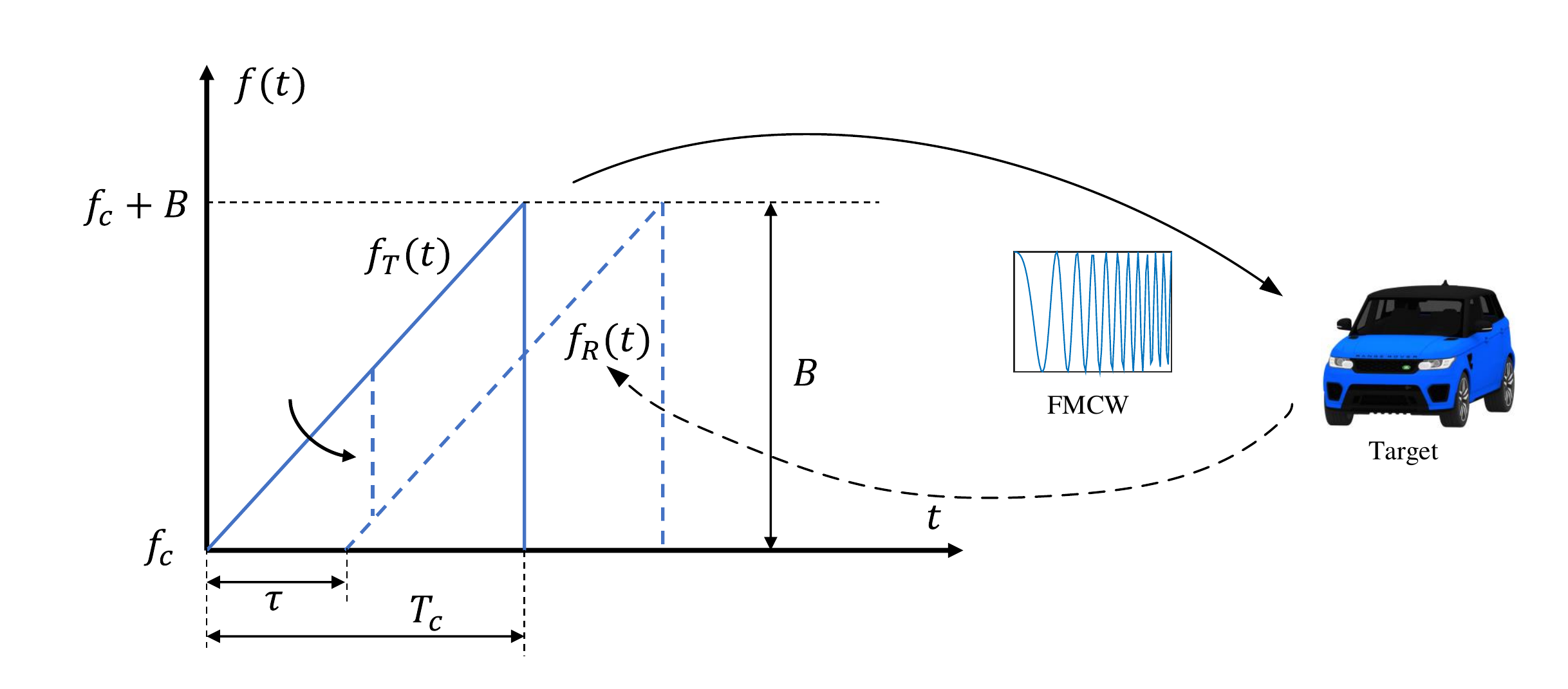}
	}
	\caption{Transmitted and received FMCW signals.}
	\label{Fig2}
\end{figure*}

As shown in Fig. \ref{Fig2}, the solid line represents the FMCW signal transmitted by the transmitting antenna, also called a chirp signal. The mathematical expression is shown in equation (\ref{1}). The dotted line represents the echo signal captured by the receiving antenna after a certain delay ${\tau}$, as shown in the equation (\ref{2}). In Fig. \ref{Fig2}, ${{\it{f}}_\text{IF}}$ represents the IF signal \cite{8930009}, ${{\it{f}}_\text{c}}$ represents the initial frequency of the FMCW, $B$ represents the frequency modulation bandwidth of the FMCW, ${{\it{T}}_\text{c}}$ represents the pulse time of the FMCW, and ${{{\mu}}={\it{B}}/{{\it{T}}_\text{c}}}$  indicates the frequency modulation slope of the FMCW.

\begin{equation}\label{1}
{{f}_{\text{T}}}\left( t \right)=\sin \left( \mathsf{2\pi }\left( {{f}_{\text{c}}}t+0.5\mu {{t}^{2}} \right)+{{\varphi }_{0}} \right)
\end{equation}
\begin{equation}\label{2}
{{f}_{\text{R}}}\left( t \right)=\sin \left( 2\mathsf{\pi }{{f}_{\text{c}}}\left( t-\tau  \right)+0.5\mu {{\left( t-\tau  \right)}^{2}}+{{\varphi }_{0}} \right)
\end{equation}

After passing through the mixer and low-pass filter, the IF signal is expressed as follows:
\begin{equation}\label{3}
{{f}_{\text{IF}}}\left( t \right)=\mathrm{LPF}\left( {{f}_{\mathrm{T}}}\left( t \right)\ast {{f}_{\mathrm{R}}}\left( t \right) \right)=\sin \left( 2\mathrm{ }\pi\left( \mu \tau t+{{f}_{\mathrm{c}}}\tau -0.5\mu {{\tau }^{2}} \right) \right)=\sin \left( 2\mathrm{ }\pi{{f}_{\mathrm{IF}}}t+{{\varphi }_{\mathrm{IF}}} \right)
\end{equation}
where $\ast$ represents the mixing operation, ${f}_{\mathrm{IF}}=\mu\tau$ represents the frequency difference between the transmitted and received signals, and ${\varphi }_{\mathrm{IF}}=2\pi {f}_{\mathsf{c}} \tau -\pi \mu{\tau}^{2}$ represents the initial phase of the IF signal, that is, the phase difference between the TX and RX when the chirp signal reaches the receiving antenna.

Fourier transform is performed on the equation (\ref{3}), and the position of the spectral peak corresponds to the IF signal ${f}_{\mathrm{IF}}$. The target distance can be obtained by the following equation (\ref{4}). When the radar transmits two or more chirp signals with a period ${{\it{T}}_\text{c}}$, the chirp signal will have a peak value at the same position after the range-FFT, but the phase is different. According to the phase difference, the velocity equation can be obtained as (\ref{5}) shown.
\begin{equation}\label{4}
r={c{{f}_\text{IF}}}/{2\mu }
\end{equation}
\begin{equation}\label{5}
v={\lambda\Delta\varphi }/{4\pi {{T}_{c}}}
\end{equation}

\begin{figure*}[!t]
	\centerline
	{
		\subfigure[]{\includegraphics[width=8cm]{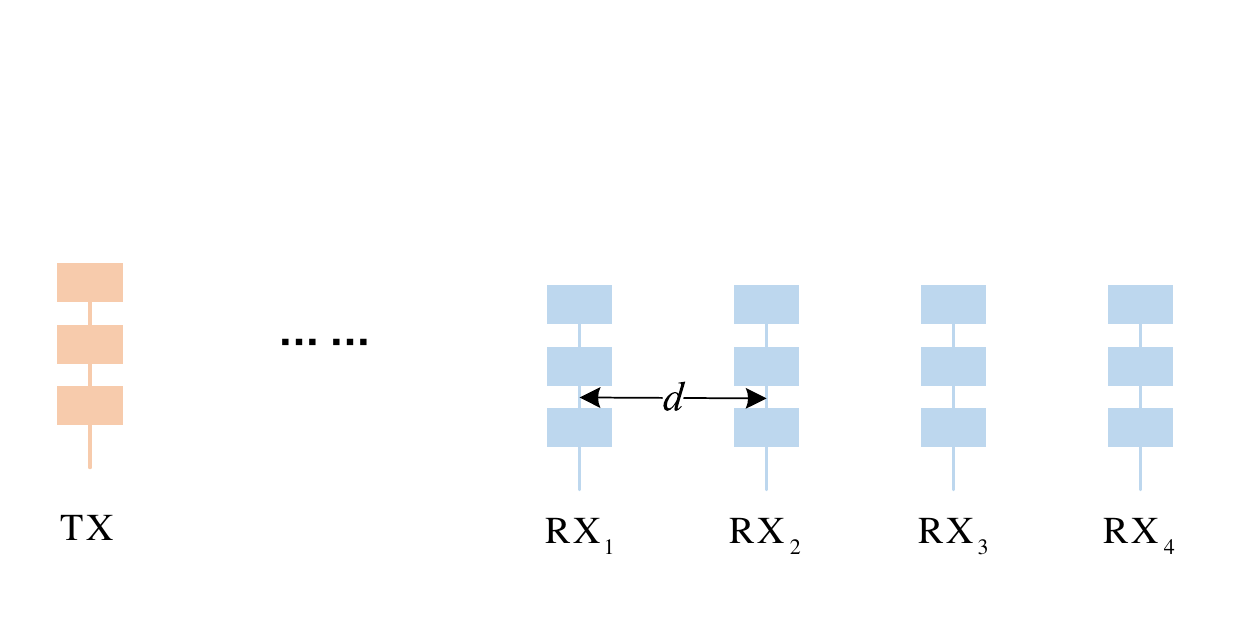}\label{Fig3a}
		}
		\hspace{.3in}
		\subfigure[]{\includegraphics[width=4cm]{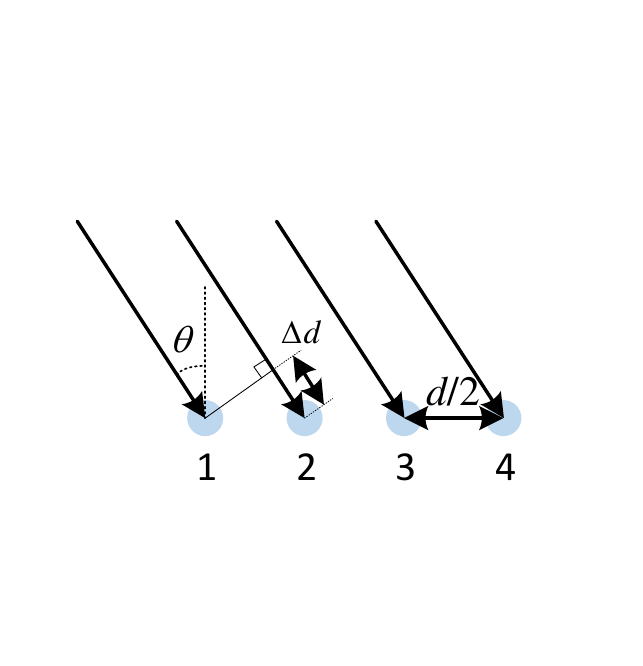}\label{Fig3b}
		}
	}
	\caption{Antenna arrays for angle estimation: (a) physical array, (b) virtual array.}
	\label{Fig3}
\end{figure*}

The angle of arrival of the radar signal can be estimated by the antenna array, as shown in Fig. \ref{Fig3a} is the 1TX4RX antenna array, and Fig. \ref{Fig3b} is the virtual array of the antenna \cite{15}. Among them, $d$ represents the physical separation distance between the two receiving antennas. Small changes in the target range also result in a phase shift of the range-FFT or Doppler-FFT peaks. When the target is in the far-field of the radar system, as shown in Fig. \ref{Fig3}, the path difference between two adjacent receiving antennas is $\Delta{\it{d}} = {\it{d}}{\sin(\theta)}$  and the angle can be calculated with $\Delta \varphi$ and  $\Delta{\it{d}}$ according to:
\begin{equation}\label{6}
\theta ={{\sin }^{-1}}\left( {\lambda \Delta \varphi }/{2\pi d}\; \right)
\end{equation}

\subsection{Radar scattering cross-sectional area}
\begin{figure*}[!t]
	\centerline
	{
		\includegraphics[width=13cm]{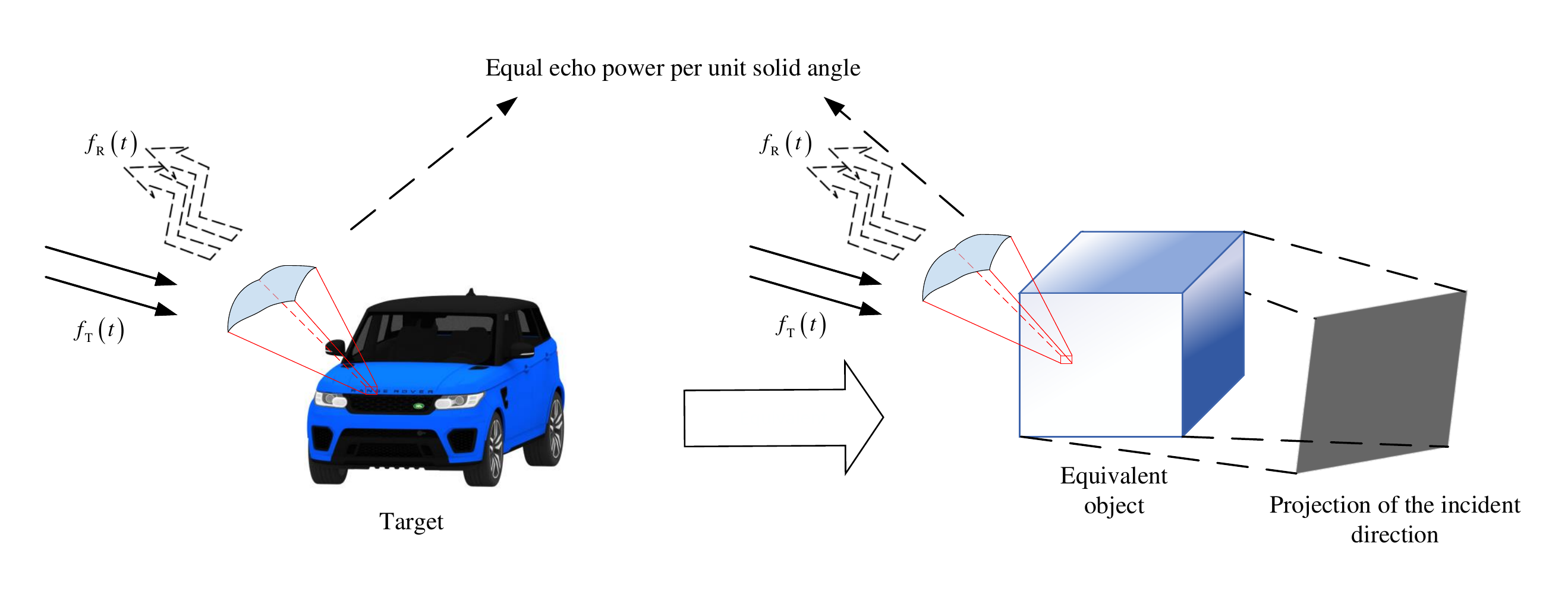}
	}
	\caption{RCS equivalent model for car target.}
	\label{Fig.4}
\end{figure*}

Radar cross section, also known as the backscatter cross section, characterizes the ability of the target to reflect the radar signal in the direction of radar reception. It is the equivalent area of the target and is represented by the projected area of a uniform equivalent reflector in the incident direction. The equivalent reflector and the target have the same echo power within the unit solid angle of the receiving direction, as shown in Fig. \ref{Fig.4}. RCS is the physical property of the target, has the same units as area, and is numerically equal to $4\pi$ times the ratio of the reflected power to the received power at the target.

When a mmw radar detects a target, the RX received power is not only related to the cross-sectional area of the target radar, but also related to the performance of the transmitter and receiver, the target distance, etc., and is commonly described as equation (\ref{7}).
\begin{equation}\label{7}
P=\frac{{{P}_{\mathrm{t}}}{{G}_{\mathrm{TX}}}\sigma {{G}_{\mathrm{RX}}}{{\lambda }^{2}}}{{{\left( 4\mathrm{\pi} \right)}^{3}}{{d}^{4}}}
\end{equation}
where $\it{P}_\text{t}$ is the output power of the device, ${\it{G}}_\text{TX},\it{G}_\text{RX}$ is the antenna gain of the transmitting and receiving antennas, $\sigma$ is the Radar Cross-Section (RCS), $\lambda$ is the wavelength of the electromagnetic wave, and ${\it{d}}$ is the distance from the target to the radar.

\begin{figure*}[!t]
	\centerline
	{
		\includegraphics[width=10cm]{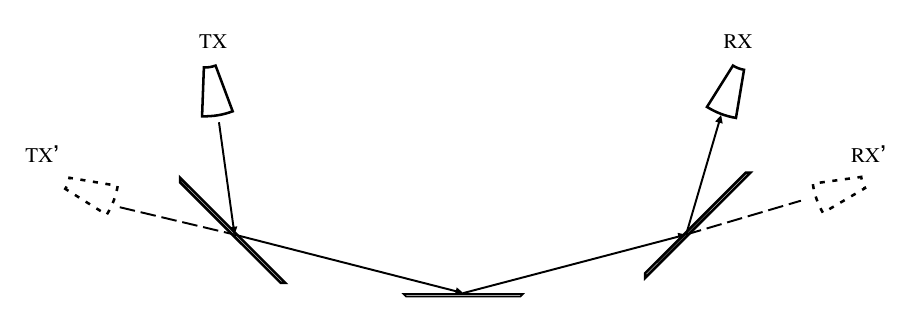}
	}
	\caption{Bidirectional analytic ray tracing diagram.}
	\label{bart}
\end{figure*}

The RCS calculation method in this paper is based on the BART algorithm proposed by F. Xu et al. \cite{16}. A ray tracing algorithm's basic idea is to emit electromagnetic waves from the antenna as many rays, keep track of each ray's scattering path until it reaches the receiver or is less powerful than a certain threshold, and then stop tracking. The BART algorithm emits rays from both the light source and the observation point, and constructs the scattering path from the source point to the observation point by connecting forward and backward rays. As shown in Fig. \ref{bart}, rays are emitted from the source (TX) in the direction of the incident and remain tracked as they are emitted to the facets and reflected. At the same time, another ray is tracked in the opposite direction of the observed scattering (RX). When the two forward and backward rays meet on the same facet, a node at that facet connects the two tracing paths, forming the scattering path from the source to the observation point. This paper uses the BART simulation tool to calculate and simulate the RCS of the target vehicle.

\section{Vehicle echo data and signal processing simulation}
To deploy mmw radar sensors at traffic intersections, it is necessary to know the optimal sensor parameters, sensor installation locations, etc., so that mmw radars can monitor traffic information more effectively and accurately. However, this requirement is unlikely to be achieved using only real-world trial and error, which requires simulation of this scenario. There are many advantages to using simulation, such as reduced time costs, variability of traffic scenarios, sensors Variability of parameters, etc. In the subsequent network training of vehicle target classification and recognition, large number of different datasets will also be required. With a reliable simulation scenario, the simulation results can be obtained easily and quickly. Next, the construction and processing flow of the radar signal of the entire simulation scenario will be introduced.
\subsection{Traffic Intersection Scenario Construction}
Mmw radar deployed at traffic intersections should face four directions and cover all lanes to supervise and control traffic information at intersections. In the simulation scenario of this paper, taking one of the road directions as an example. The radar is set on the traffic light pole at the intersection. The Cartesian coordinate system is established with the projection point of the radar sensor on the ground as the origin O and assumes that the lane direction detected by radar is Y-direction, vertical ground upward is z-direction, and vertical to the Y-O-Z plane is X-direction. The following content in this paper follows this coordinate system setting. The specific parameters of the simulated scenario are shown in Table \ref{tab:table1}.
\begin{table}[htb]
	\caption{Simulation scenario parameters\label{tab:table1}}
	\centering
	\begin{tabular}{l c c}
		\toprule
		\text{ } & Corresponding coordinate axis & Parameters (unit: meters) \\
		\midrule
		Radar altitude   & Z-axis  &  6 \\
		Intersection width & Y-axis  & 22.5 \\
		Lane width  & X-axis & 3.75 \\
		First lane center point  & X-axis & 2 \\
		Second lane center point  & X-axis & 5.75\\
		Third lane center point  & X-axis & 9.5 \\
		\bottomrule
	\end{tabular}
\end{table}

The target vehicle randomly appears in these three lanes, approaching the traffic intersection at a random speed under the premise of being realistic. The speed, distance, lane, and other traffic information of the target vehicle should be accurately and efficiently obtained.

\subsection{Vehicle Echo Generation}
In this section, the simulation of the FMCW echo signal will be introduced, and it is mainly for the vehicle target single frame signal data modeling calculation. The method proposed in this paper is simulated based on the entire link of the mmw radar detection target system, and its working principle is shown in Fig. \ref{Fig1}. The system transmits the FMCW signal, and its mathematical expression is described as follows:
\begin{equation}\label{8}
{{x}_{\mathrm{T}}}(t)=\exp \left\{ \mathrm{j}2\mathrm{\pi}\left( {{f}_{0}}t+0.5\mu {{t}^{2}} \right)+{{\varphi }_{0}} \right\}
\end{equation}
where ${\it{f}_{0}}$ is the initial frequency of the transmitted signal, $\mu$ is the slope of the linear change of the frequency of the transmitted signal, and ${\varphi_{0}}$ is the initial phase of the transmitted signal.

\begin{figure*}[!t]
	\centerline
	{
		\subfigure[]{\includegraphics[width=4.5cm]{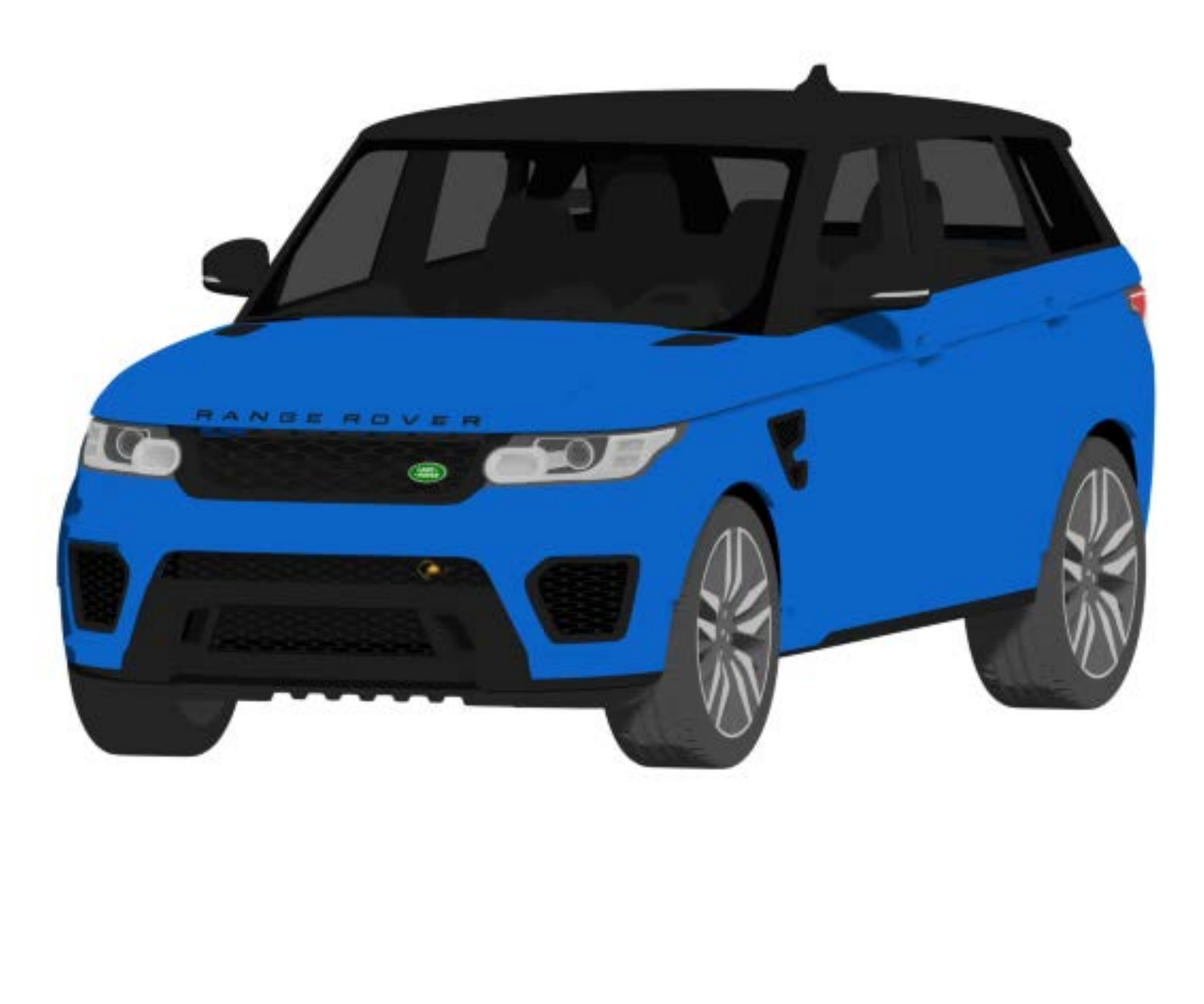}\label{Fig4a}
		}
		\hspace{.3in}
		\subfigure[]{\includegraphics[width=6cm]{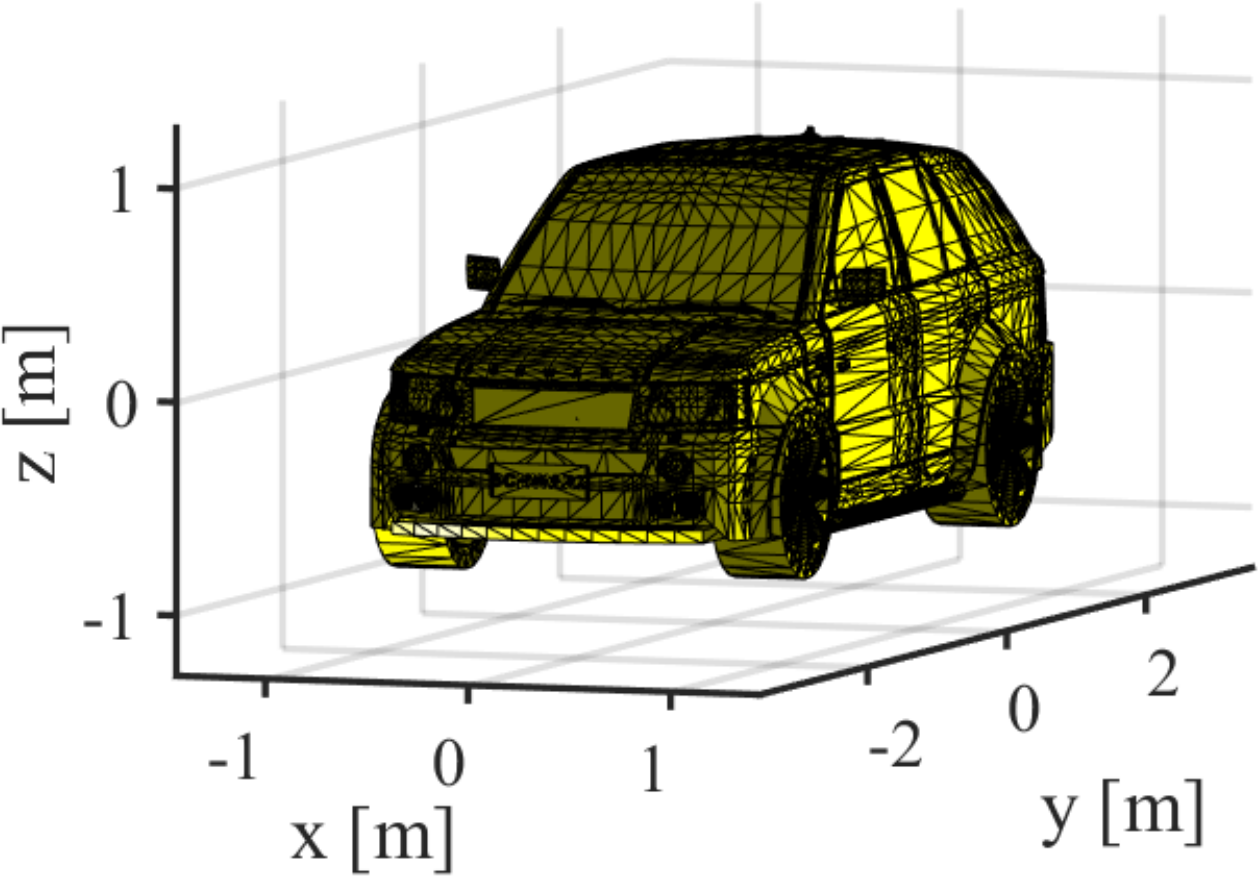}\label{Fig4b}
		}
	}
	\caption{Land Rover vehicle 3D model: (a) The original model; (b)The model processed by MATLAB.}
	\label{Fig4}
\end{figure*}

When the transmitted signal encounters the target vehicle, part of the energy will be captured and reflected by the receiving antenna. During this process, the radiation intensity of the reflected signal from the target in the direction of radar reception is determined by the RCS. The simulation calculation requires a digital model to represent a car in the real physical world, such as the vehicle model used in Fig. \ref{Fig4} (Land Rover's Range Rover), which is about 5 meters long, 2 meters wide, and 1.8 meters high.

In this paper, the RCS of the target vehicle is obtained by using the BART simulation tool introduced in Section 2. By calculating the position of the target at all chirp times in a frame, the azimuth and pitch angles between the target and the radar are obtained. Combining the digital model of the target and the sweep frequency range input to BART, the frequency response ${\sigma}_{\text{f}}$ of the target RCS is obtained. Combined with the radar equation of equation (\ref{7}), the frequency domain expression of the echo signal can be calculated with:
\begin{equation}\label{9}
{{x}_{\mathrm{R}}}\left( f \right)=P\cdot {{\sigma }_{\mathrm{f}}}\otimes {{x}_{\mathrm{T}}}\left( f \right)
\end{equation}
where $\otimes$ represents the convolution and ${\it{P}}$ is the energy at the receiving antenna. To simplify the computational complexity, it is converted from a convolution operation in the frequency domain to a product in the time domain by an inverse Fourier transform. The transformation equation can be expressed as follows:
\begin{equation}\label{10}
\mathrm{IFFT}\left[ {{x}_{\mathrm{R}}}\left( f \right) \right]=\mathrm{IFFT}\left[ P\cdot {{\sigma }_{\mathrm{f}}}\otimes {{x}_{\mathrm{T}}}\left( f \right) \right]
\end{equation}
\begin{equation}\label{11}
{{x}_\mathrm{R}}\left( t \right)=P\cdot {{\sigma }_{\mathrm{t}}}\cdot \exp \left\{ {\mathrm{j2\pi}} \left[ {{f}_{0}}\left( t-\tau \left( t \right) \right)+0.5\mu {{\left( t-\tau \left( t \right) \right)}^{2}} \right]+{{\varphi }_{0}} \right\}
\end{equation}
where  ${\sigma }_{\mathrm{t}}$ represents the radar scattering cross-sectional area, ${\it{\tau}(t)}$ is the round-trip time of the radar signal to detect the target, which can be calculated by the following equation:
\begin{equation}\label{12}
\tau \left( t \right)=2\left( d+vt \right)/c
\end{equation}
where $d$ is the initial distance from the vehicle to the radar, $v$ is the speed of the vehicle, and $\text{c}$ is the speed of light. After receiving the echo signal and mixing it with the transmitting signal, the IF signal can be obtained, and its expression is
\begin{equation}\label{13}
{{x}_{\mathrm{IF}}}\left( t \right)=P\cdot {{\sigma }_{\mathrm{t}}}\exp \left\{ {\mathrm{j2\pi}} \left[ {{f}_{0}}\tau \left( t \right)+\mu t\tau \left( t \right)-0.5\mu \tau {{\left( t \right)}^{2}} \right] \right\}
\end{equation}

To simulate the environmental noise existing in the real scenario, Gaussian white noise with a certain signal-to-noise ratio can be added to the simulation process.

\subsection{Radar Signal Processing}

\begin{figure*}[!t]
	\centerline
	{
		\includegraphics[width=\linewidth]{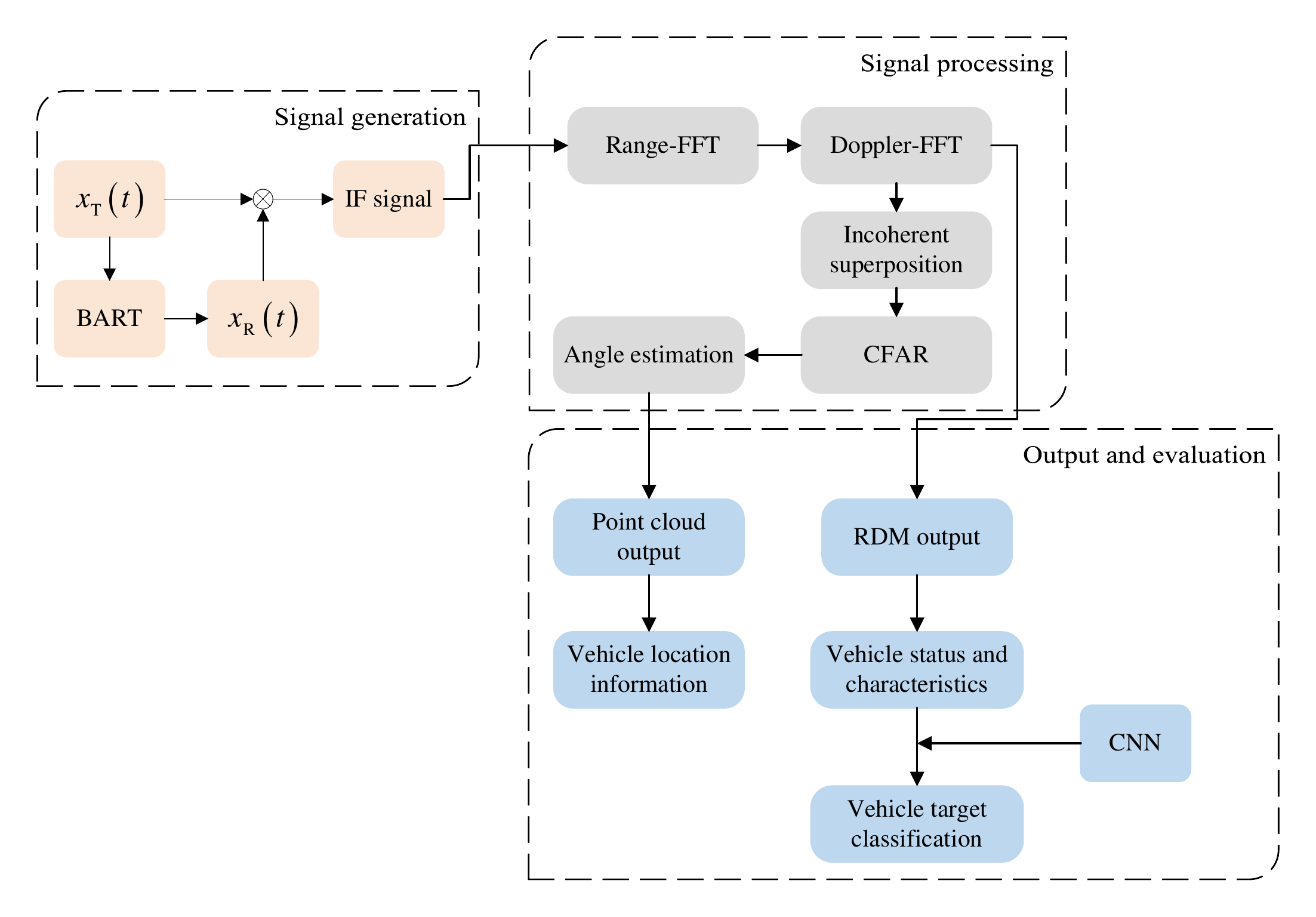}
	}
	\caption{Flowchart of radar signal simulation.}
	\label{Fig5}
\end{figure*}

The simulation flow of radar signal processing in this paper is shown in Fig. \ref{Fig5}. In each frame, the transmitting antenna transmits N chirps, and each chirp has M sampling points. The radar system has P transmitting antennas and Q receiving antennas, with a total of O (P×Q) channels, resulting in an O×M×N data matrix. The RDM output is obtained by the fast Fourier transform of the range dimension and Doppler dimension. Then, the constant false alarm rate detection is carried out, using the CA-CFAR algorithm \cite{17}, setting the appropriate threshold, effectively eliminating the influence caused by the interference noise, and finding the target point of interest. Then the target angle information can be obtained by performing the fast Fourier transform on the angle dimension of each channel. Combined with the range and angle information, the position information of the target point relative to the radar can be obtained, so that the point cloud image can be output \cite{18}.

\subsection{Dataset and convolutional neural network}
\begin{figure*}[!t]
	\centerline
	{
		\subfigure[]{\includegraphics[width=4cm]{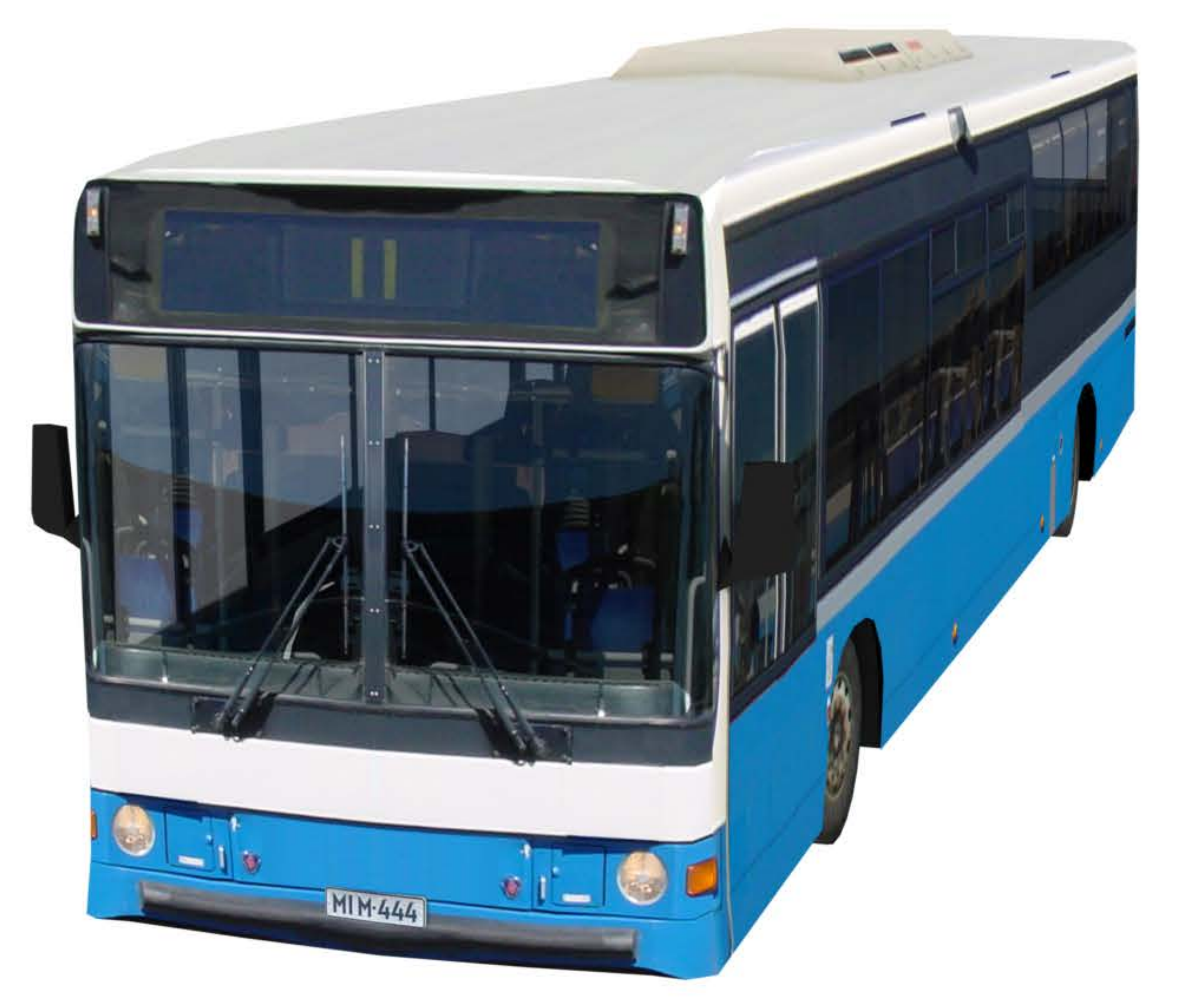}\label{Fig6a}
		}
		\hspace{.5in}
		\subfigure[]{\includegraphics[width=4cm]{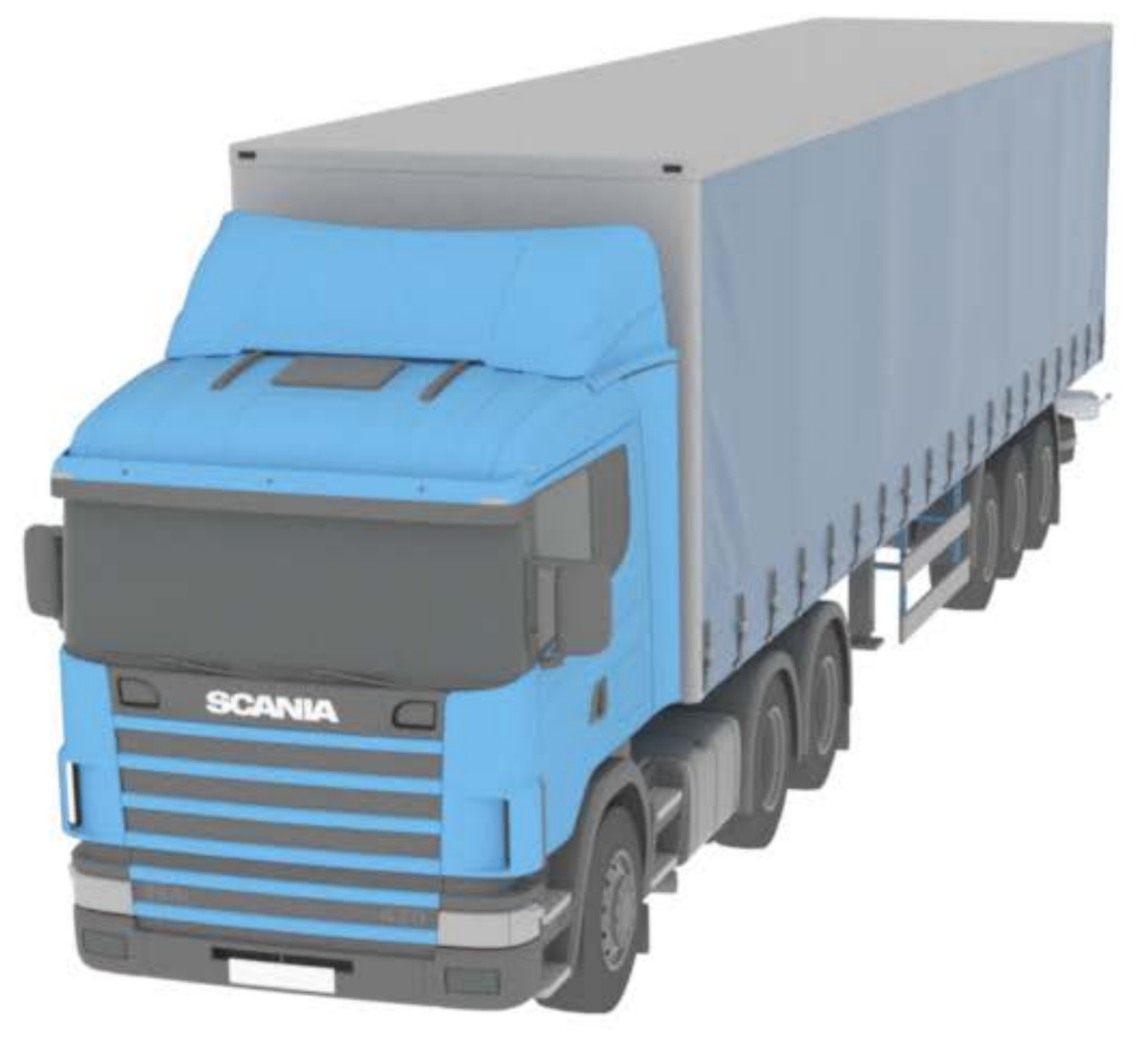}\label{Fig6b}
		}
		\hspace{.5in}
		\subfigure[]{\includegraphics[width=4cm]{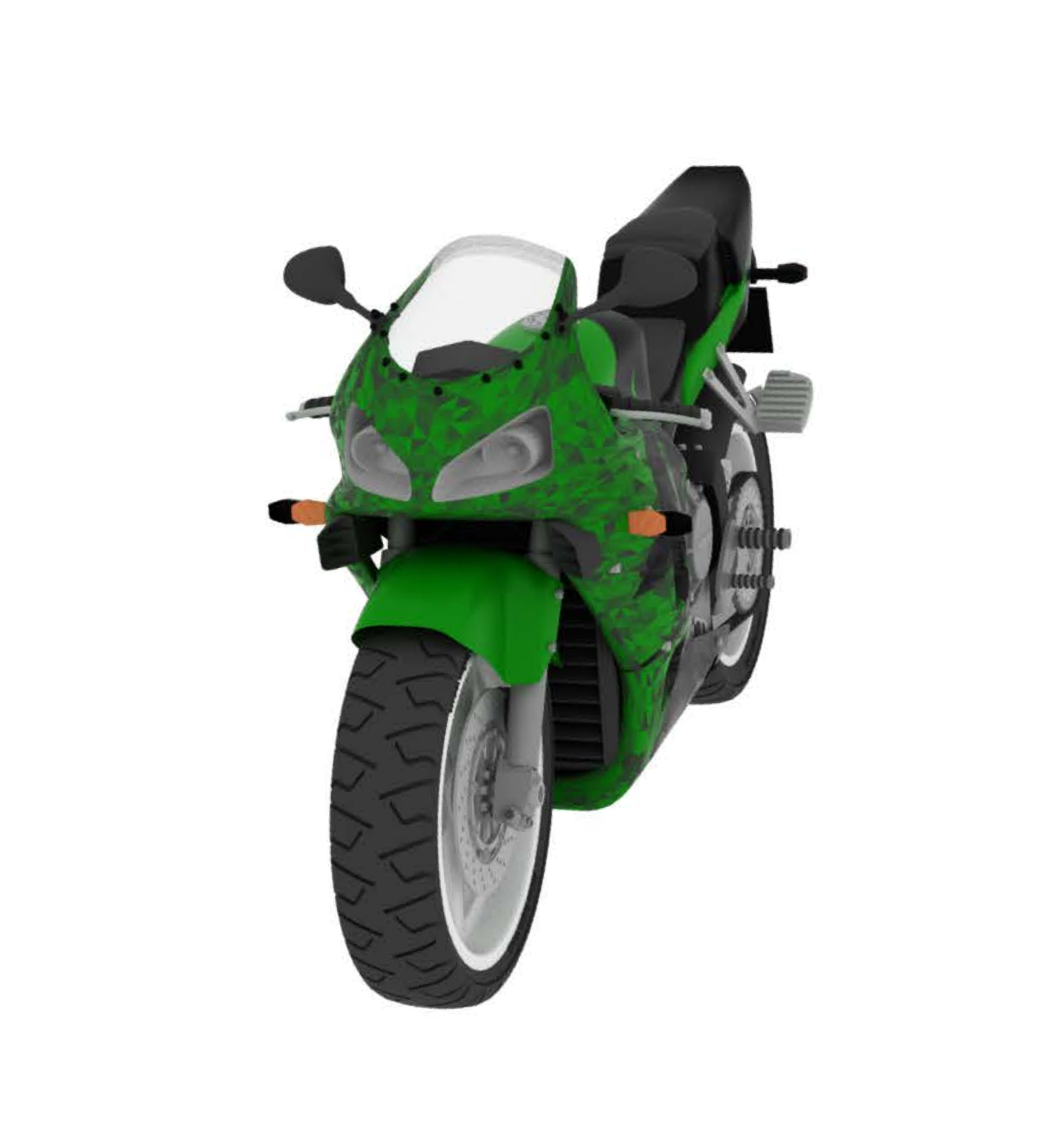}\label{Fig6c}
		}
	}
	\caption{Vehicle original models: (a) bus; (b) truck; (c) motorcycle.}
	\label{Fig6}
\end{figure*}

To verify the availability of the data of the radar simulation method and analyze the accuracy of the classification and recognition of the vehicle target by the mmw radar, a convolutional neural network is designed in this paper. The target type is discriminated based on the RDM of the detection output of the mmw radar. This paper mainly studies four types of vehicle objects involved in transportation, namely cars, buses, trucks and motorcycles. The model shown in Fig. \ref{Fig4} is used in the car class, and the vehicle models in the other categories are shown in Fig. \ref{Fig6}. 

Put these four types of vehicle models into the simulation scenario introduced in Section 3, and constantly change the distance (25 m $\sim$ 40 m), velocity (0.5 m/s $\sim$ 10 m/s) and the lanes (1, 2, 3). The four models were tested 20 times and 80 RDMs were obtained.

Afterward, a series of preprocessing is performed on the data set. The original output color RDM is converted into grayscale images to reduce the calculation amount of the classification network. The interesting parts are extracted from 80 existing images, and the data set is expanded by image operations such as translation, rotation, adding Gaussian noise, and adding salt and pepper noise. Finally, 280 images of each type of vehicle were obtained, and a total of 1120 images were used as the data set of the subsequent training classification network. 

\begin{figure*}[!t]
	\centerline
	{
		\includegraphics[width=12cm]{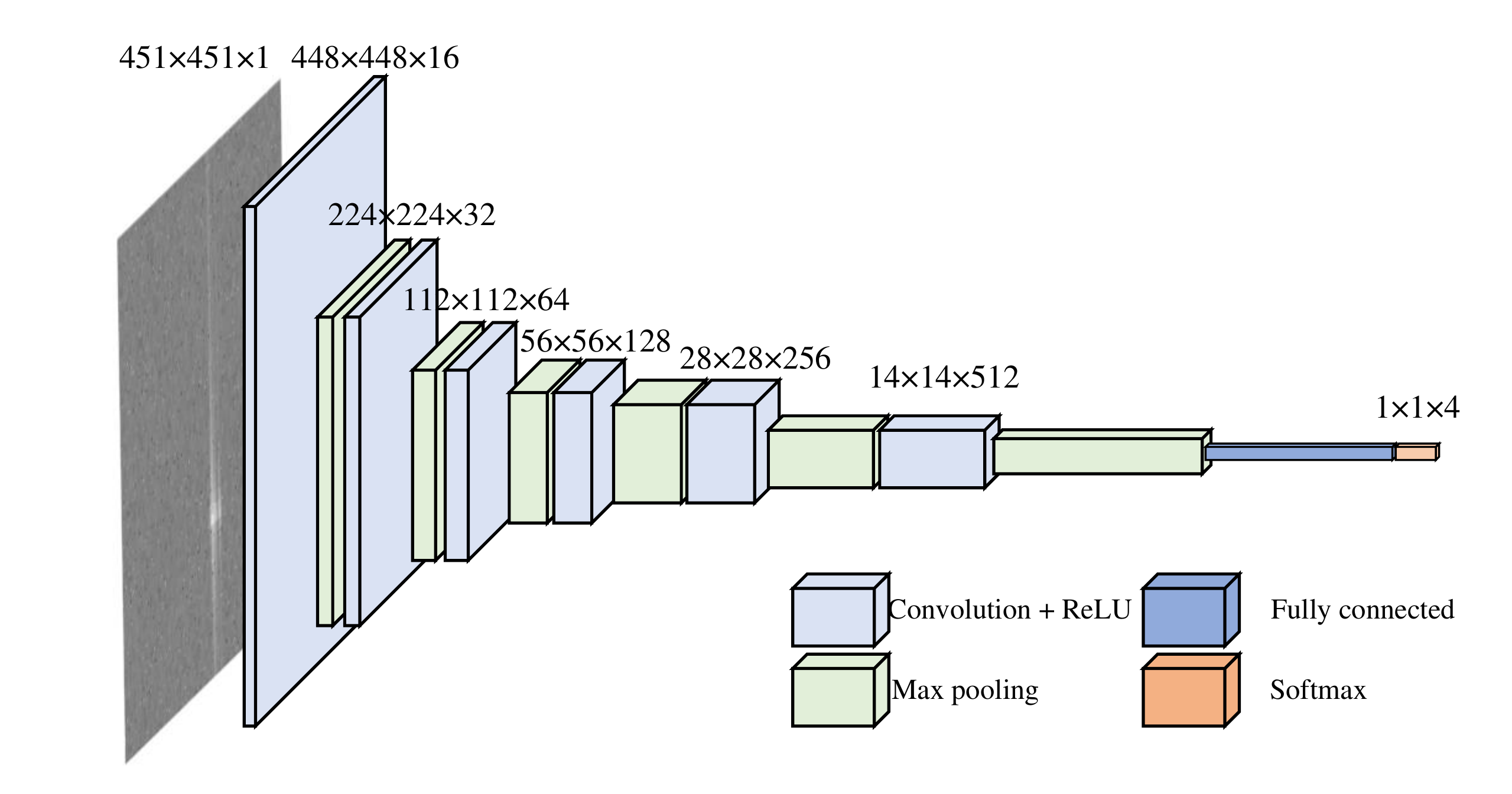}
	}
	\caption{The model structure of the classification network.}
	\label{Fig7}
\end{figure*}

The model structure of the classification network is shown in Fig. \ref{Fig7}, and the input size of the network is the size of dataset 451×451×1. The convolution layer in the network structure includes three parts: the convolution layer, the normalization layer, and the linear activation function. The pooling layer uses the maximum pooling down sampling, and the fully connected layer outputs the probability of each category. 

\section{Experiment and analysis}
\subsection{Experiment setup}
To sample without distortion, the sampling rate must be greater than the maximum IF signal, in this experiment it is 5.12 Msps. Combined with the actual traffic intersection situation, the preset requires the maximum detection distance of the radar to be 60 m, and the frequency slope of the FMCW is 12.5 $\text{MHz}/\mu\text{s}$. From the limitation of radar detection range resolution, a higher range resolution means larger bandwidth. When the frequency modulation slope and the sampling rate are fixed, the sampling time can be increased by increasing the number of sampling points to meet the requirements of increasing the bandwidth and improving the distance resolution. However, when the sampling time is increased, the maximum speed that can be measured by the radar sensor will be reduced due to the limitation of the maximum speed detected by the radar. In the trade-off between range resolution and maximum speed, several sets of experiments were carried out in this paper, the number of sampling points of a single chirp of the radar sensor was adjusted, and the results were simply compared. The experimental parameters and some results are shown in Table \ref{tab:table2}.

\begin{table}[htb]
	\caption{Simulated experimental data under different parameters\label{tab:table2}}
	\centering
	\begin{tabular}{p{2in} p{1in} p{1in} p{1in}}
		\toprule
		Experiment number & 1 & 2 & 3 \\
		\midrule
		Target vehicle distance   & 40 m  &  40 m & 40 m \\
		Target vehicle speed & 10 m/s & 10 m/s & 10 m/s \\
		Sampling points  & 256 & 386 & 512\\
		Pulse time  & 60 $\mu$s & 85 $\mu$s & 110 $\mu$s \\
		Bandwidth  & 0.625 GHz & 0.9375 GHz & 1.25 GHz\\
		Distance resolution  & 0.24 m & 0.16 m & 0.12 m \\
		Measurement speed (Max)  & 16.16 m/s & 11.38 m/s & 8.78 m/s \\
		RDM display  & normal & normal & abnormal \\
		Point cloud points  & 14 & 15 & 18 \\
		\bottomrule
	\end{tabular}
\end{table}

\begin{figure}[htbp]
	\centering
	\subfigure[RDM(256 sampling points)]{
		\includegraphics[width=0.45\linewidth]{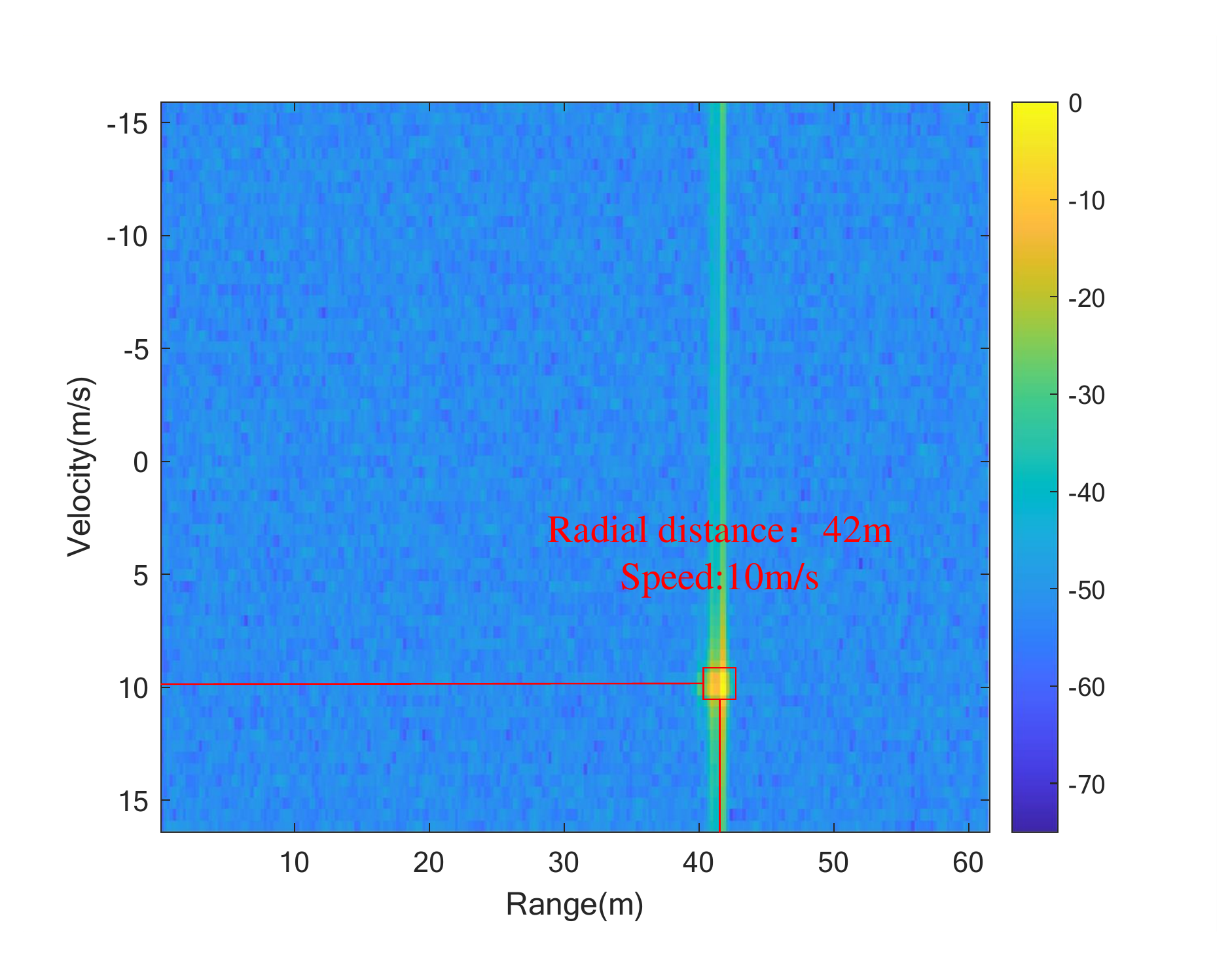}
		\label{figrdm-pc-a}
	}
	\subfigure[Point cloud(256 sampling points)]{
		\includegraphics[width=0.45\linewidth]{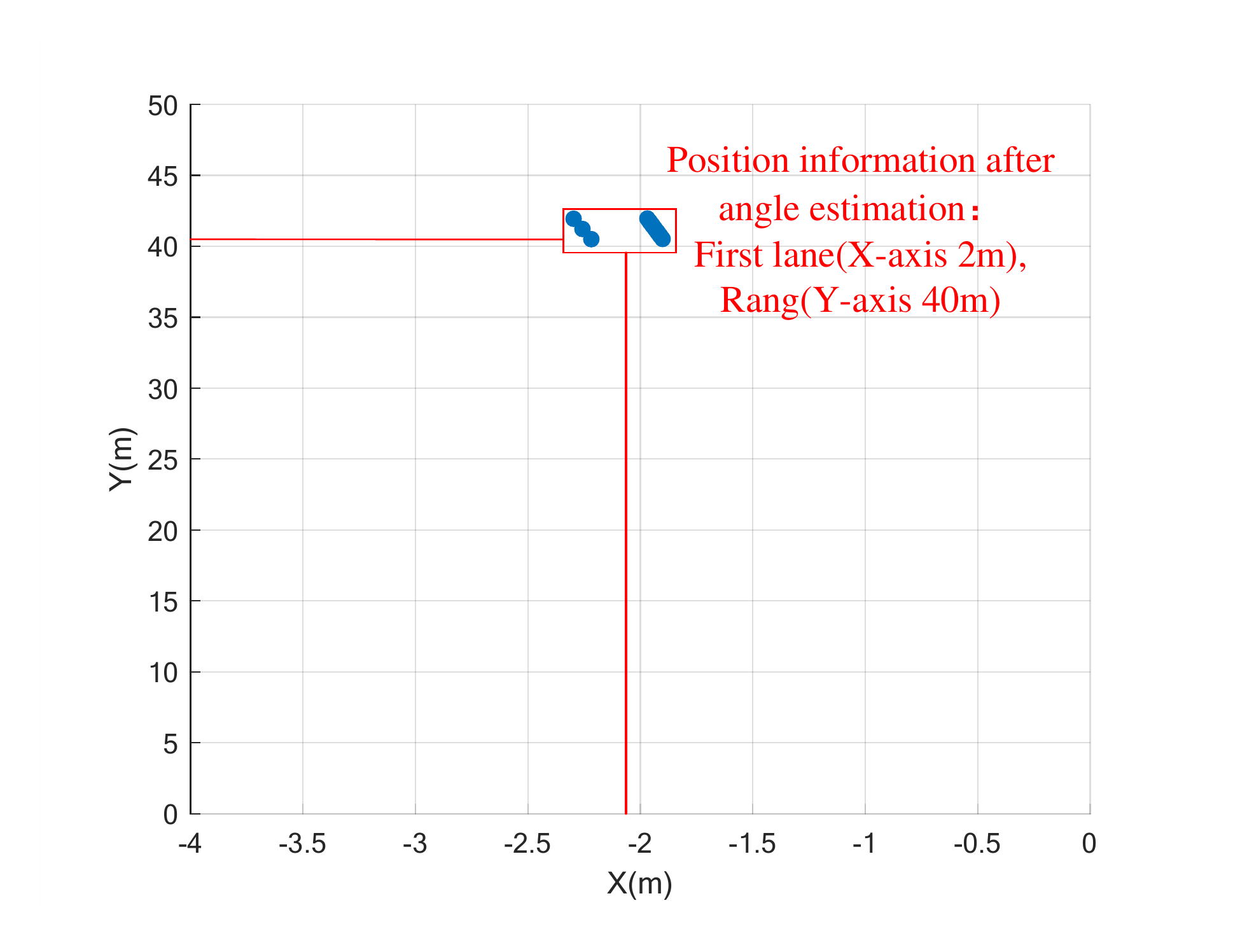}
		\label{figrdm-pc-d}
	}
	\quad 
	\subfigure[RDM(384 sampling points)]{
		\includegraphics[width=0.45\linewidth]{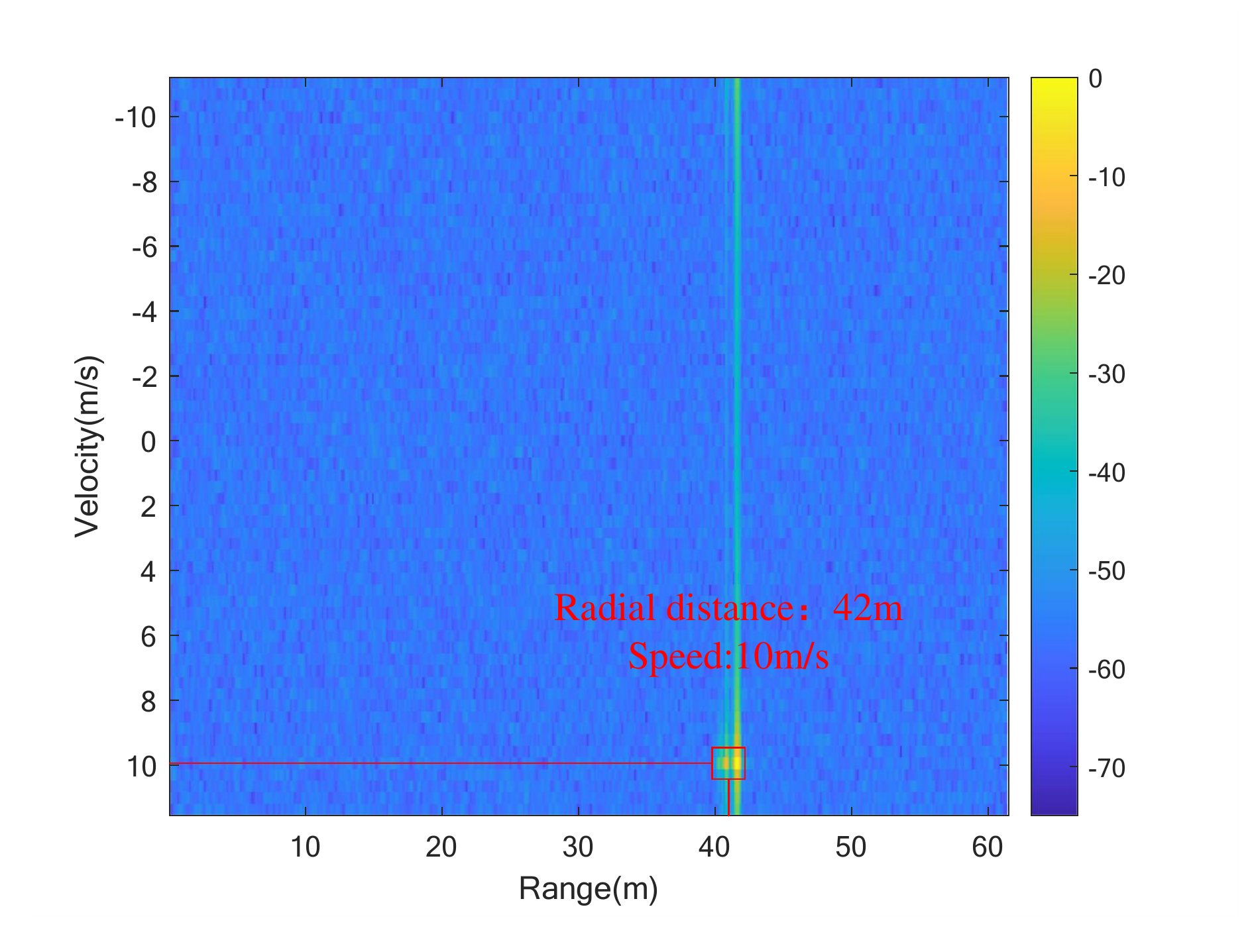}
		\label{figrdm-pc-b}
	}
	\subfigure[Point cloud(384 sampling points)]{
		\includegraphics[width=0.45\linewidth]{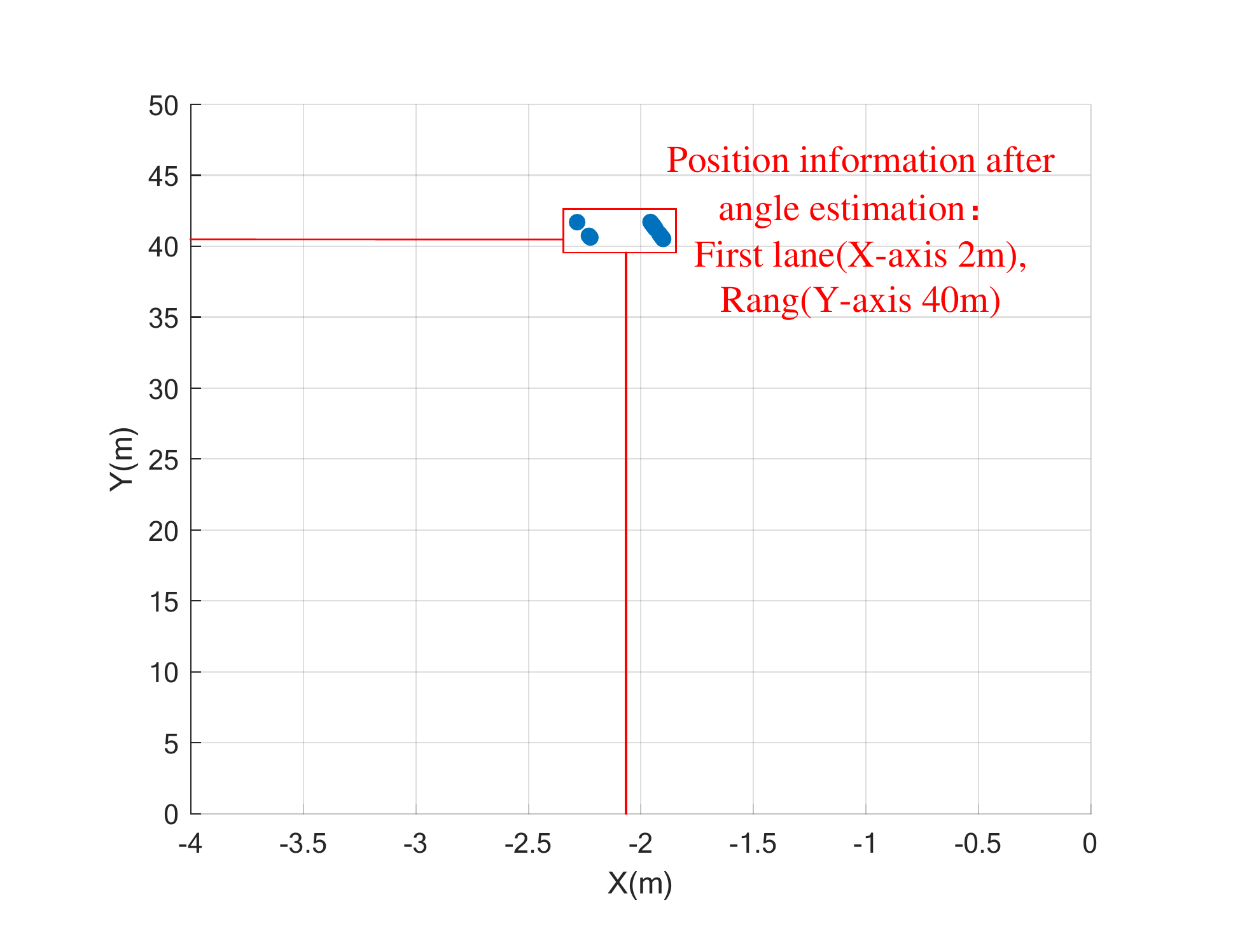}
		\label{figrdm-pc-e}
	}
	\quad  
	\subfigure[RDM(512 sampling points)]{
		\includegraphics[width=0.45\linewidth]{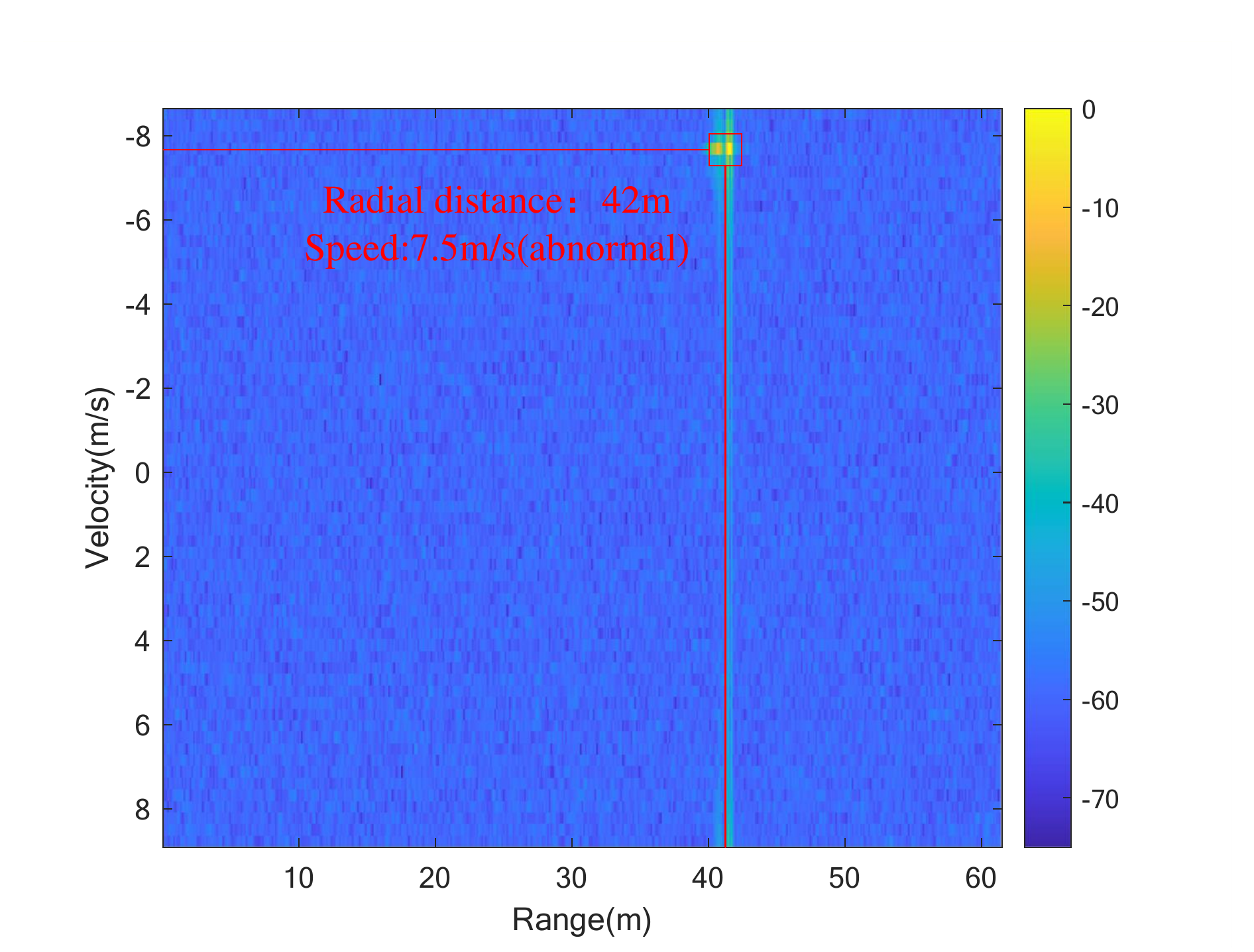}
		\label{figrdm-pc-c}
	}
	\subfigure[Point cloud(512 sampling points)]{
		\includegraphics[width=0.45\linewidth]{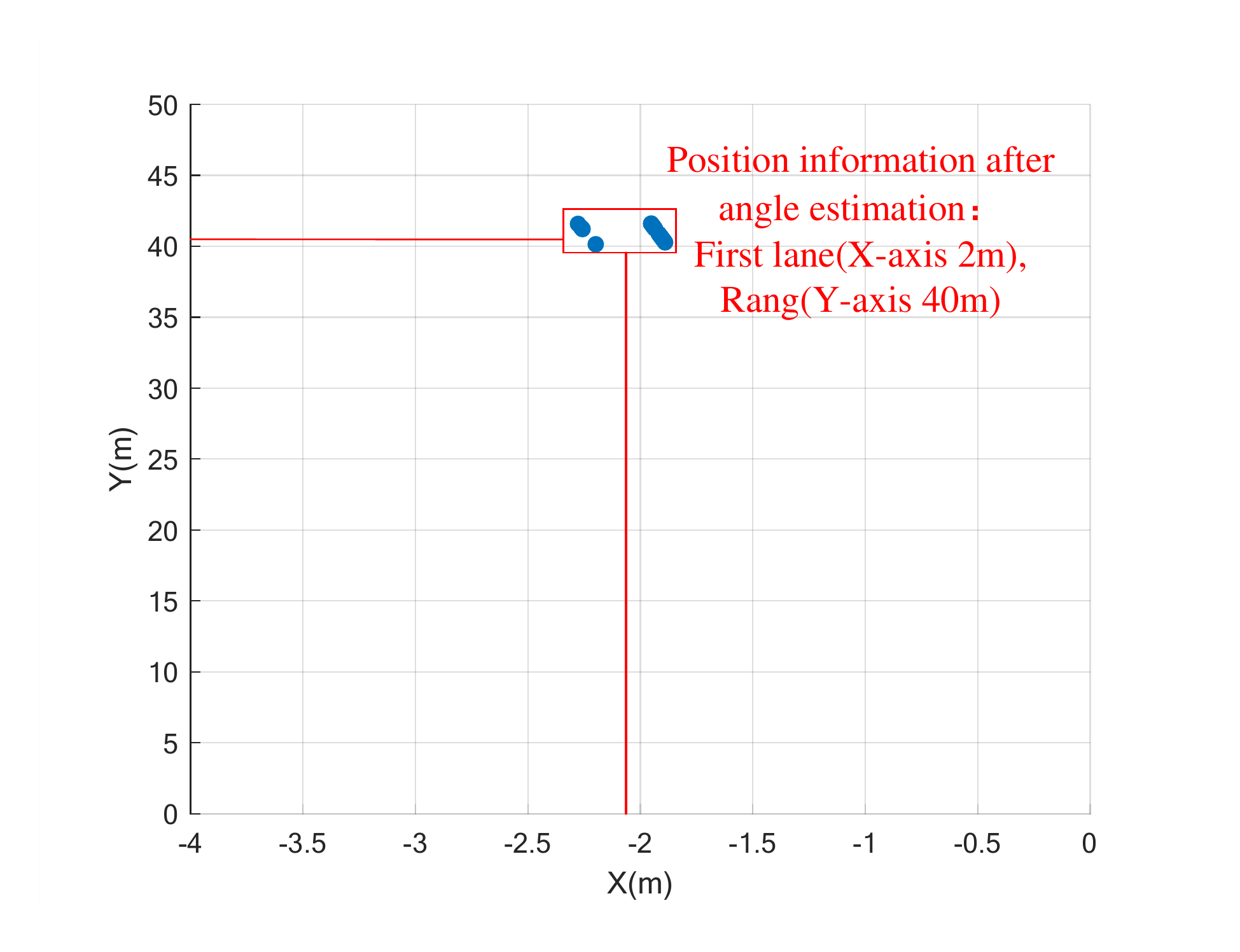}
		\label{figrdm-pc-f}
	}
	\caption{The output results of the three experiments described by Land Rover in Table \ref{tab:table2}, the RDM contains distance and speed information, and the point cloud contains position information}
	\label{figrdm-pc}
\end{figure}

\subsection{RDM and point cloud analysis}
The output results of the three experiments described in Table \ref{tab:table2} are shown in Fig. \ref{figrdm-pc}, which separately show the RDM output and the point cloud image output.

In the RDM image, the peak point represents the identified target vehicle, the abscissa represents the distance, and the ordinate represents the speed. The coordinate (0, 0) point in the output of the point cloud image is the location of the radar, and the point cloud represents the position of the vehicle target. Combining Table \ref{tab:table2} and Fig. \ref{figrdm-pc}, when the speed of the target vehicle exceeds the maximum speed detectable by the radar, although the point cloud image output is still normal, the speed display in the RDM is abnormal. The experimental results between 256 and 384 sampling points are not much different, but when the number of sampling points is set to 384, the sampling time of a single frame is prolonged, and the amount of calculation increases. Under a similar effect, the sensor sampling point was set to 256 for subsequent experiments.

The above experiments show that the deployment of the mmw radar in practical scenarios is limited by hardware such as radar sensor sampling rate and IF signal frequency. To improve the range resolution, while increasing the bandwidth, it is also necessary to consider meeting the needs of vehicle speed detection. Based on the sensor parameters described in this paper and large number of experiments, vehicle targets traveling at speeds ranging from 0 to 15 m/s (0 to 54 Km/h) and at distances ranging from 0 to 60 m can be accurately identified.

\subsection{Classification and recognition performance analysis}
\begin{figure}[!t]
	\centering
	\subfigure[Car]{
		\includegraphics[width=0.45\linewidth]{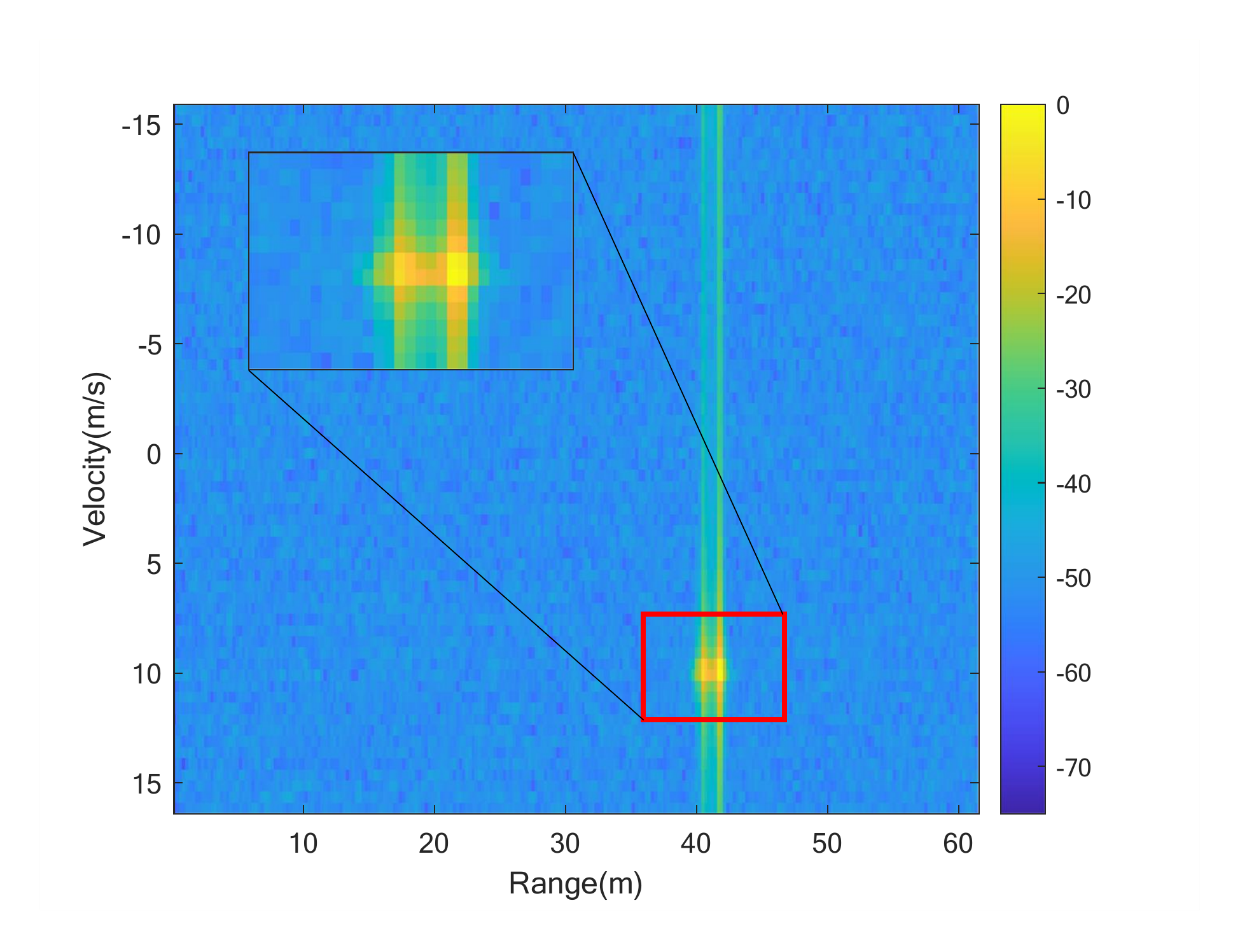}
		\label{figrdlh}
	}
	\subfigure[Bus]{
		\includegraphics[width=0.45\linewidth]{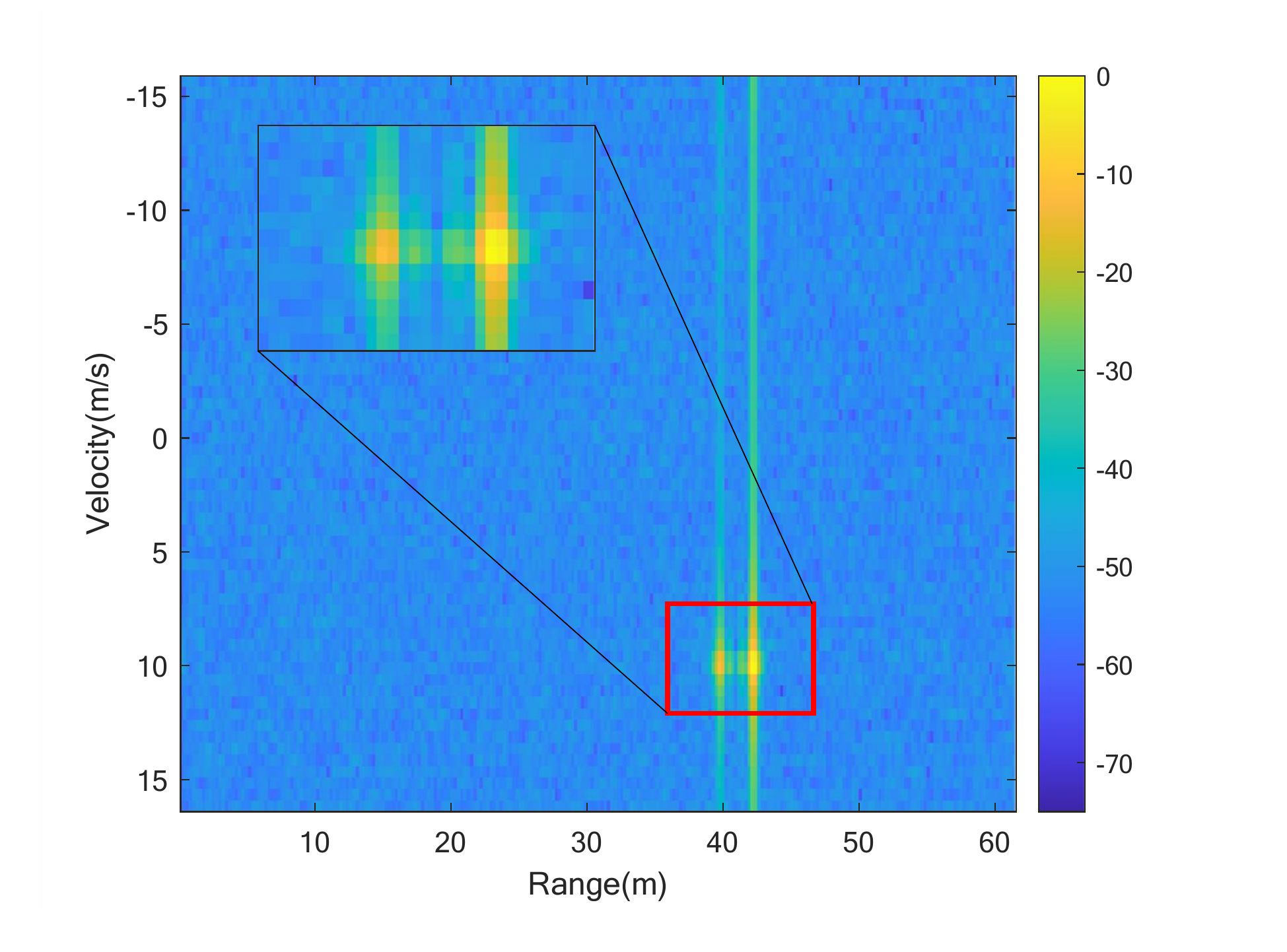}
		\label{figrdbus}
	}
	\quad 
	\subfigure[Truck]{
		\includegraphics[width=0.45\linewidth]{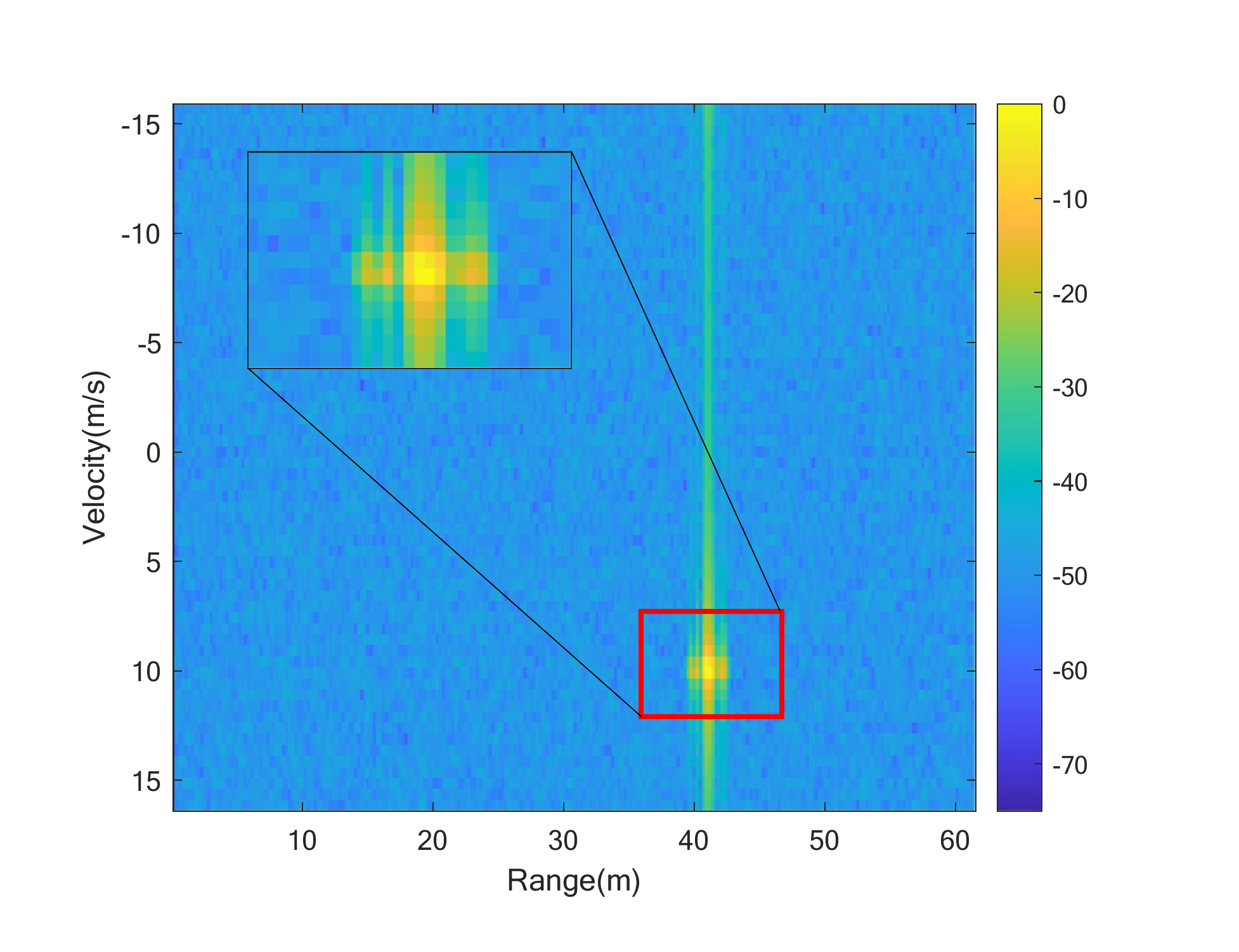}
		\label{figrdtruck}
	}
	\subfigure[Motorcycle]{
		\includegraphics[width=0.45\linewidth]{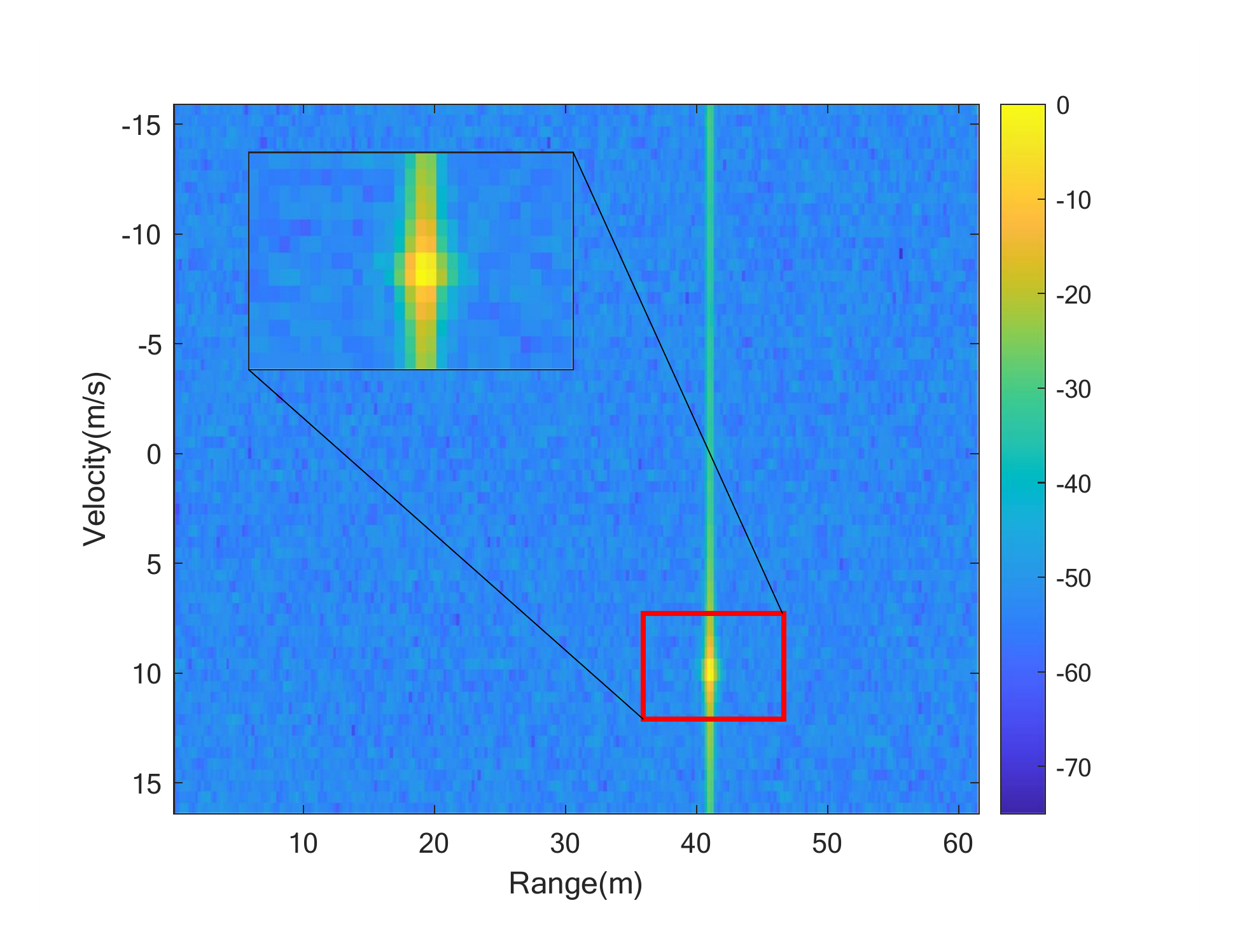}
		\label{figrdmotorcycle}
	}
	\caption{Differences in RDM output of four types of vehicle models under the same experimental parameters.}
	\label{figrd}
\end{figure}

The RMD output contains the state and feature information of the target. Under the same parameters, the RDM outputs of the four types of vehicles used in this paper are shown in Fig. \ref{figrd}.

\begin{figure*}[!t]
	\centerline
	{
		\includegraphics[width=0.95\linewidth]{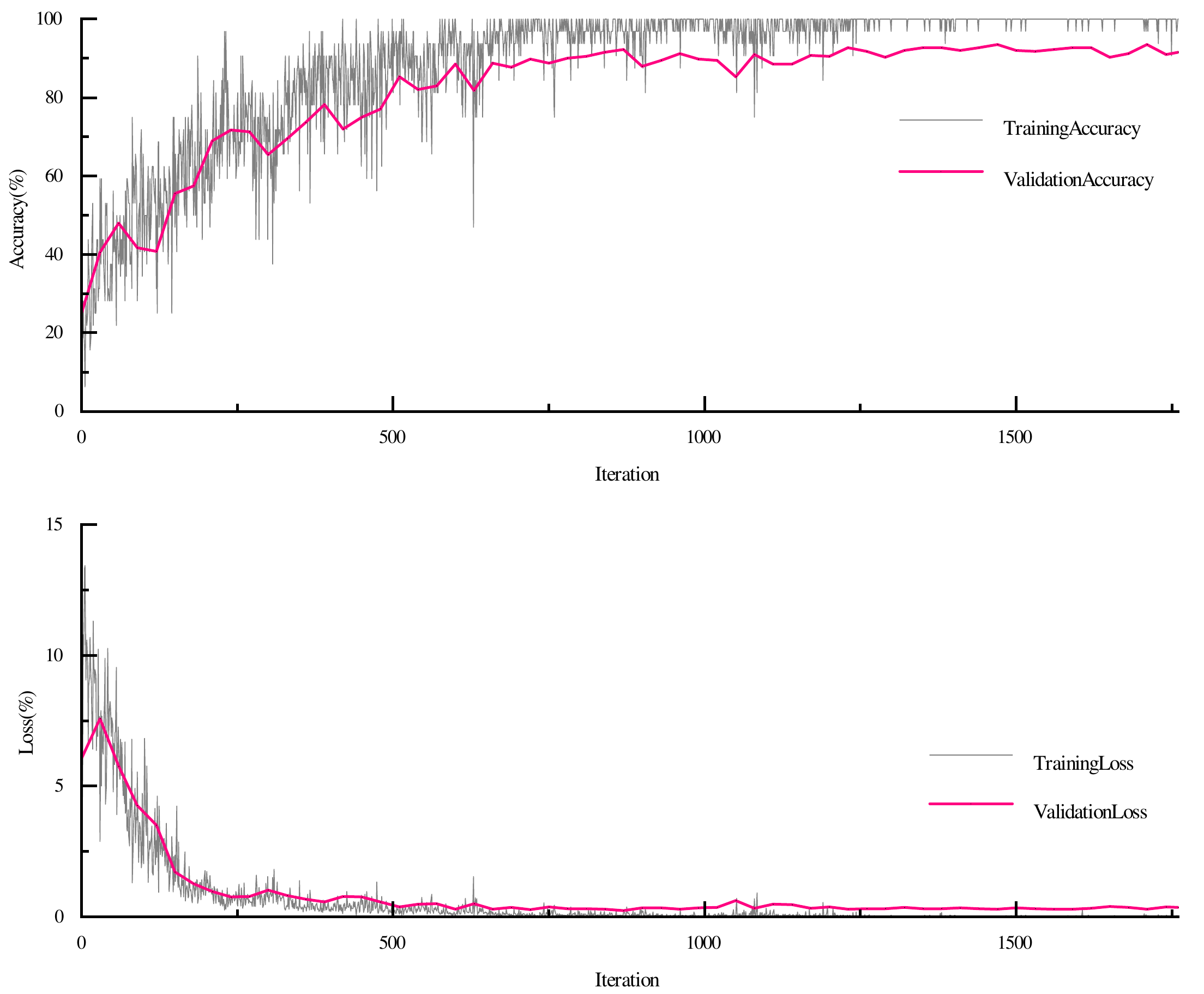}
	}
	\caption{Classification network training accuracy curve (top), loss rate curve (bottom).}
	\label{Fig11}
\end{figure*}

\begin{figure*}[!t]
	\centerline
	{
		\includegraphics[width=0.7\linewidth]{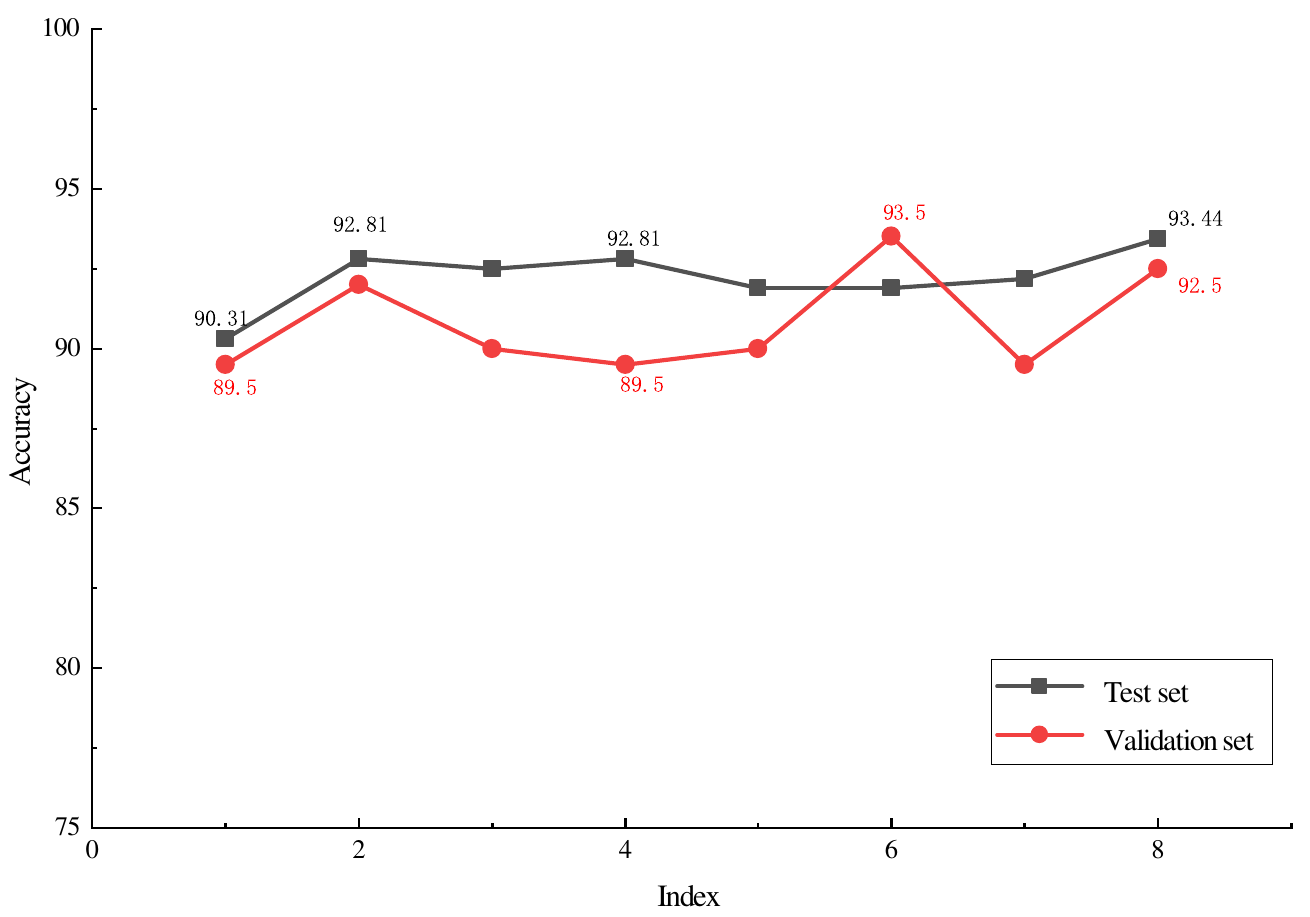}
	}
	\caption{The accuracy of the classification network on the test set and the validation set (containing 200 new images).}
	\label{Fig12}
\end{figure*}

In the process of training the network, 1120 images were randomly scrambled and divided into a training set containing 600 images, a test set containing 320 images, and a validation set containing 200 images. The training curve of the classification network is shown in Fig. \ref{Fig11}.

Using the training set and test set above, the learning rate was set at 0.005, and after 1440 iterations, the training of the classification network was completed. Network performance was then tested using 200 unused images. Fig. \ref{Fig12} shows some of the results.

\begin{figure*}[!t]
	\centerline
	{
		\includegraphics[width=\linewidth]{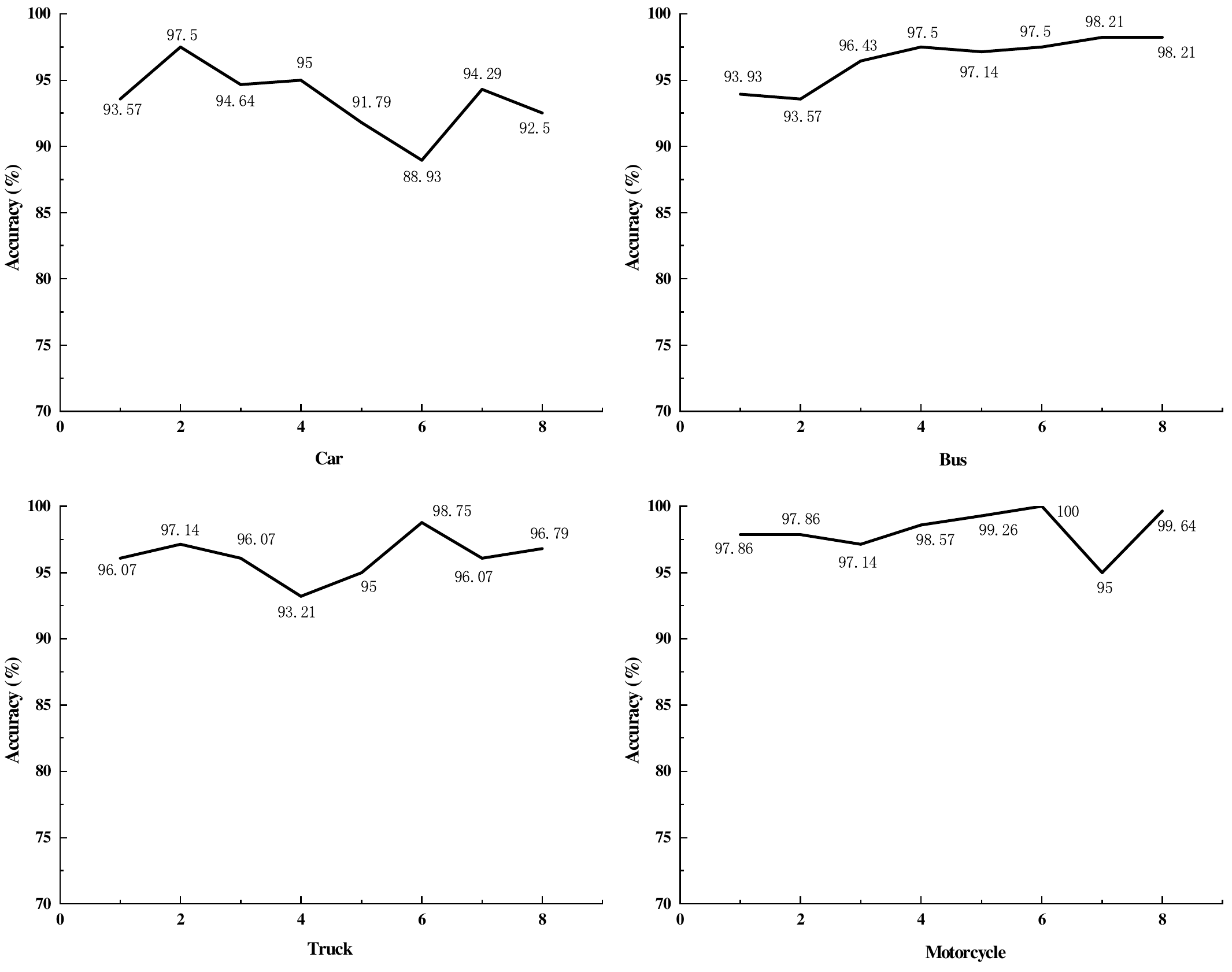}
	}
	\caption{Classification accuracy of four types of vehicles.}
	\label{Fig13}
\end{figure*}

After continuous iterative learning, the average accuracy of the classification network on the 320 images in the test set reached 92.23\%. Although the accuracy generally dropped when classifying the 200 images in the validation set, the average accuracy was still 90.81\%. In this paper, the four types of vehicles used in the experiment are separately classified and verified, which are divided into cars, buses, trucks, and motorcycles and some of the results are shown in Fig. \ref{Fig13}.

The average classification accuracy rate of buses and trucks is about 96\%, and the classification accuracy rate of cars is also about 93.5\%. Because motorcycles have more obvious characteristics than other vehicles, the classification accuracy rate is also the highest, with an average of 98.16\%.

\section{Conclusion}
This paper proposes a simulation method for mmw radar detection and recognition in traffic intersection applications. In this method, a simulated traffic intersection scenario is constructed, and the method of mmw radar echo generation is studied, especially combining it with BART, and its signal processing process is considered. By using this method, it is possible to provide useful and accurate data when deploying mmw radars at actual intersections. This helps to expand the application research of mmw radar in the field of transportation, such as vehicle trajectory tracking, vehicle violation discrimination, etc. It provides theoretical guidance and analysis tools for the future development of mmw radar systems that comply with relevant frequency bands. The method in this paper is a preliminary attempt based on roadside sensor simulation, which basically meets the needs of intersection applications and realizes target detection and recognition. In the future, we can try to combine this method with existing tools, such as CARLA, to achieve real-time scene simulation.

\label{}





\bibliography{mybibfile}





\end{document}